\documentclass[aps,prx,showpacs,showkeys,amssymb,amsmath,superscriptaddress,reprint]{revtex4-2}

\usepackage{graphicx,bm,amsmath}
\usepackage[usenames,dvipsnames,table]{xcolor}
\usepackage[colorlinks,bookmarks=false,citecolor=NavyBlue,linkcolor=Red,urlcolor=blue,
]{hyperref}
\usepackage{xr-hyper}      
\usepackage{appendix}

\usepackage[bbgreekl]{mathbbol}
\usepackage[normalem]{ulem}
\usepackage{braket}
\usepackage{comment}
\usepackage{amsthm}
\usepackage{enumerate}

\usepackage{diagbox}
\usepackage{mdframed}
\usepackage[table]{xcolor}
\usepackage{colortbl}
\usepackage{array}
\def\doi{http://dx.doi.org/}
\newcommand{\be}{\begin{equation}}
\newcommand{\ee}{\end{equation}}
\newcommand{\bec}{\begin{equation*}}
\newcommand{\eec}{\end{equation*}}
\newcommand{\bea}{\begin{eqnarray}}
\newcommand{\eea}{\end{eqnarray}}

\usepackage{orcidlink}

\begin{document}
\title{Topological magic response in quantum spin chains}
\author{Ritu Nehra~\orcidlink{0000-0002-8983-2588}}
\affiliation{The Abdus Salam International Centre for Theoretical Physics (ICTP), Strada Costiera 11, 34151 Trieste, Italy}
\affiliation{Dipartimento di Fisica e Astronomia “G. Galilei”, Università di Padova, I-35131 Padova, Italy}

\author{Poetri Sonya Tarabunga~\orcidlink{0000-0001-8079-9040}}
\affiliation{Technical University of Munich, TUM School of Natural Sciences,
Physics Department, 85748 Garching, Germany}
\affiliation{Munich Center for Quantum Science and Technology (MCQST), Schellingstr. 4, 80799 M\"unchen, Germany}

\author{Martina Frau}
\affiliation{International School for Advanced Studies (SISSA), via Bonomea 265, 34136 Trieste, Italy}

\author{Mario Collura~\orcidlink{0000-0003-2615-8140}}
\affiliation{International School for Advanced Studies (SISSA), via Bonomea 265, 34136 Trieste, Italy}
\affiliation{INFN Sezione di Trieste, 34136 Trieste, Italy}

\author{Emanuele Tirrito~\orcidlink{0000-0001-7067-1203}}
\email{emanuele.tirrito@epfl.ch}
\affiliation{Laboratory of Theoretical Physics of Nanosystems (LTPN), Institute of Physics, Ecole Polytechnique Fédérale de Lausanne (EPFL), CH-1015 Lausanne, Switzerland}
\affiliation{Center for Quantum Science and Engineering, Ecole Polytechnique Fédérale de Lausanne (EPFL), CH-1015 Lausanne, Switzerland}

\author{Marcello Dalmonte~\orcidlink{0000-0001-5338-4181}}
\affiliation{The Abdus Salam International Centre for Theoretical Physics (ICTP), Strada Costiera 11, 34151 Trieste, Italy}
\affiliation{Dipartimento di Fisica e Astronomia, Università di Bologna, via Irnerio 46, I-40126 Bologna, Italy}

\begin{abstract}
Topological matter provides natural platforms for robust, non-local information storage, central to quantum error correction. Yet, while the relation between entanglement and topology is well established, little is known about the role of nonstabilizerness (or magic), a pivotal concept in fault-tolerant quantum computation, in topological phases. We introduce the concept of topological magic response, the ability of a state to spread over stabilizer space when perturbed by finite-depth non-Clifford circuits. 
Unlike a topological invariant or order parameter, this response function probes how a phase reacts to non-Clifford perturbations, revealing the presence of non-local quantum correlations.
In Ising-type spin chains, we show that symmetry-broken and paramagnetic phases lack such a response, whereas symmetry-protected topological (SPT) phases always display it. To capture this, we utilize a combination of stabilizer R\'{e}nyi entropies that, in analogy with topological entanglement entropy, isolates non-locally stored information. Using exact analytic computations and matrix product states simulations based on an algorithmic technique we introduce, we show that SPT phases doped with $T$ gates support robust topological magic response, while trivial phases remain featureless.

\end{abstract}

\maketitle
\section{Introduction}
Topological phases of matter are central to both condensed matter physics and quantum information~\cite{nielsen00}. Their hallmark is the presence of long-range entanglement and non-local correlations that remain robust against local perturbations~\cite{RevModPhys.80.1083}, providing natural platforms for fault-tolerant quantum memories~\cite{KITAEV20032,PhysRevLett.102.110502}, which are now actively pursued in experiments. 
The interplay between entanglement and topology has been extensively studied~\cite{zeng2015,Zeng2016,moessner2021topological,Nehra2025controlling} - most notably through the concept of topological entanglement entropy~\cite{kitaev2006, levin2006} and entanglement spectra~\cite{li2008entanglement}.

From a quantum computational perspective, fault tolerance warranted by topology alone does not guarantee universality: achieving it requires going beyond Clifford operations~\cite{gottesman1997stabilizer,Gottesman1998,Gottesman98theory,Aaronson04}. The essential additional resource is nonstabilizerness — or magic — which quantifies a state’s departure from the stabilizer framework, and sets the ultimate cost for fault tolerant universal quantum computation based on stabilizer codes. While, over the last few years, there has been a growing effort in understanding the role of nonstabilizerness in relation to in- and out-of-equilibrium collective phenomena~\cite{tarabunga23m, Frau24a,10.21468/SciPostPhys.18.5.165, 
tarabunga2023magic, turkeshi2025magic,tarabunga2024critical,PhysRevA.106.042426,rz86-47h3,1jzy-sk9r,y9r6-dx7p,PhysRevB.111.054301,p8dn-glcw,magni2025quantum,xfp5-hhs4}, its connection to topological effects is presently not clear at all, neither qualitatively nor quantitatively. 
\begin{figure}[h!]
    \centering
   \includegraphics[width=\linewidth]{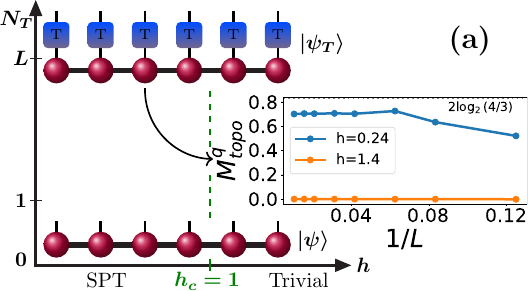}\\
   \includegraphics[width=\linewidth]{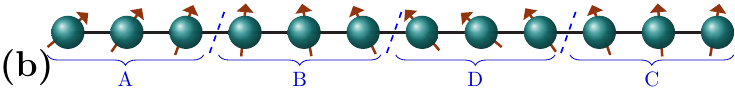}
    \caption{(a) Schematic of the ground state $|\psi\rangle$ and the T-gate–doped state $|\psi_T\rangle$ with $N_T$ number of T-gates. The transverse field $h$ tunes between the symmetry-protected topological (SPT) and trivial phases, with the transition occurring at the critical point $h = h_c$. The topological magic ($M^q_{\rm topo}$) response, shown at the center, is nonzero in the SPT phase and vanishes in the trivial phase as the system size $L$ varies.
    (b) Spin-chain partitioning into four segments: $A = [1,L/4], B = (L/4, L/2], D = (L/2, 3L/4]$, and $C = (3L/4, L]$.}
    \label{fig:intro}
\end{figure}
In fact, several paramagnetic as well as topological phases are described by stabilizer states, giving no clear hint on the role of magic on the latter, and in fact signalling, if anything, the {\it absence} of a direct connection~\cite{ellison2021symmetry,catalano2025quantum}.

In this work, we close this gap introducing the notion of topological magic response: the ability of a quantum phase to encode non-local information in terms of how much spread the state becomes in stabilizer space when perturbed by finite-depth non-Clifford circuits.  
This operational concept captures how a phase can “host” distributed magic resources. While conventional phases such as symmetry-broken or paramagnetic states do not typically exhibit a non-local magic response, we demonstrate that symmetry-protected topological (SPT) phases generically support it. 

To formalize this idea, we define topological stabilizer R\'{e}nyi Entropy (TSRE) in direct analogy with topological entanglement entropy: just as the latter isolates the universal non-local contribution to entanglement beyond short-range correlations, topological magic extracts the non-local component of nonstabilizerness, quantified in terms of participation entropy on stabilizer space - stabilizer Rényi entropies~\cite{Leone2022stabilizer}. 
This refinement is motivated by recent results showing that the total magic is unable to detect the topological character of certain phases~\cite{catalano2025quantum}, even if it can still be non-trivially bounded with respect to paramagnets~\cite{PRXQuantum.3.020333}. By contrast, TSRE succeeds in distinguishing topologically trivial phases, where magic remains local, from those supporting intrinsically global nonstabilizerness correlations. Crucially, TSRE shall be measured after the application of a constant-depth magic circuit made of local unitaries, which serves as a way to generalize states within the same phase: this is the core of magic response, and is a {\it sine qua non} condition to circumvent fine tuned scenarios.

Concretely, as illustrated in Fig.~\ref{fig:intro}, we consider ground states of quantum spin chains, and probe their magic response by injecting a finite number $N_T$ of non-Clifford $T$ gates. Panel (a) of Fig.~\ref{fig:intro} shows the schematic construction: starting from a ground state $|\psi\rangle$, we apply $N_T$ $T$ gates to obtain the doped state $|\psi_T\rangle$, and study how the resulting magic depends on the underlying phase of matter and on the tuning parameter $h$ that drives the transition between trivial and SPT phases. We quantify non-local contributions via the TSRE $M^{q}_{\mathrm{topo}}$, extracted from the quadri-partition scheme illustrated in Fig.~\ref{fig:intro}(b). 

Using a combination of analytic calculations and large-scale Pauli-MPS simulations~\cite{tarabunga24a}, we show that trivial and symmetry-broken phases exhibit only local, additive magic with vanishing topological contributions, while SPT phases robustly support non-local topological magic. In particular, the magic response of SPT ground states doped with $T$ gates is universal: it yields finite values of $M^{q}_{\mathrm{topo}}$ that closely parallel topological entanglement entropy, identifying SPT phases as natural hosts of intrinsically non-local magic resources - whose exact amount depends on the circuit details. By contrast, conventional phases remain topologically trivial under the same perturbations.

Our approach connects to the concept of long-range SRE~\cite{PhysRevB.103.075145,tarabunga23m,sarkar2020characterization}, as well as operational ways of quantifying long-range magic~\cite{ellison2021symmetry,wei2025long,korbany2025long,4brj-cl26,ahmad2025experimentaldemonstrationnonlocalmagic}. The relevance of the former is by now clear in the context of conformal field theories theories~\cite{PhysRevB.103.075145,tarabunga23m,hoshino2025stabilizer,hoshino2025stabilizer2,rajabpour2025stabilizer} (including relations to measures of magic ~\cite{tarabunga2024critical,sarkar2020characterization,hoshino2025stabilizer,timsina2025robustness}), state complexity~\cite{10.21468/SciPostPhys.18.5.165,Frau24a} and quantum dynamics~\cite{tarabunga2024magic,lopez2024exact}. It reveals that magic, like entanglement, can display universal non-local features, but is a strictly stronger resource, linking quantum error correction, computational complexity, and many-body physics. The latter is instead directly tied to the operational capabilities of quantum circuits, of direct relevance to fault-tolerant quantum computing architectures. As we describe below, the TSRE we define is very tightly related to long-range SRE; however, while conceptually close, a direct relation to long-range magic is not fully developed.

\section{Results}

\subsection{Quantifying nonstabilizerness}

To quantify the spreading of a given state on the stabilizer set, we utilize the stabilizer R\'{e}nyi entropy (SRE)~\cite{Leone2022stabilizer}, defined for a state $|\psi\rangle$ in terms of Pauli strings on $L$ qubits $\mathcal{P}_L=\left\lbrace P_1 \otimes P_2 \cdots \otimes P_L\right\rbrace_{P_j=I,X,Y,Z} $ as
\begin{equation} \label{eq:sre1}
M_n(|\psi\rangle) = \frac{1}{1-n}\log_2\left[\sum_{P\in \mathcal{P}_{L}}\frac{\langle \psi|P|\psi \rangle^{2n}}{2^L}  \right],
\end{equation}
where $L$ is the number of qubits, $n$ is the R\'enyi index, and $P$ is a Pauli string that belongs to the Pauli group $\mathcal{P}_L$. In particular, $M_1$ is defined by the limit $n\to 1$ in~\eqref{eq:sre1}, and $M_{n}\ge 0$, with the equality holding if and only if $|\Psi\rangle$ is a stabilizer state~\cite{haug2023stabilizerentropiesand,gross2021schurweylduality}. 
Moreover, the definition of SREs can be extended to mixed states by properly normalizing it~\cite{tarabunga2025efficient}:
\begin{align}
\tilde{M}^K_n(|\psi\rangle)=\frac{1}{1-n}\log_2\Bigg\{\frac{\displaystyle\sum_{P\in\mathcal{P}^\prime_{L_K}} |\text{Tr}( \rho_K P)|^{2n}}{\displaystyle\sum_{P\in\mathcal{P}^\prime_{L_K}}|\text{Tr}\rho_K P_K|^2}\Bigg\},
\label{Mix_SRE}
\end{align}
where the denominator accounts for state purity. For subsystems $K$, $\tilde{M}^K_n$ is computed over the reduced Pauli group $\mathcal{P}^\prime_{L_K}$, consisting of Pauli strings acting only on the $L_K$ sites of $K$ (identity elsewhere). This corresponds to the SRE of the reduced density matrix $\rho_K$, capturing the local or subsystem magic. While the mixed-state SREs are not proper measures of magic for certain Hilbert space dimensions, they can be used to construct mixed-state magic witnesses~\cite{haug2025efficient}. Our work stresses, however, their relevance as a participation entropy in stabilizer space, whose strong physical significance has already been established in conformal field theory~\cite{hoshino2025stabilizer,tarabunga23m}.
In our study, we fix $n=2$ for both pure and mixed state SREs.

One advantage of the SRE over many other proposed measures of magic~\cite{Veitch14} is that it allows an efficient computation even for a large $L$ ~\cite{PhysRevB.107.035148,tarabunga23m,lami23a,tarabunga24a,Frau24a,PhysRevB.111.085144,ding2025evaluating}.
Inspired by these advances, we define a related quantity, topological stabilizer R\'{e}nyi Entropy (TSRE), defined as:
\begin{align}
M^{q}_{\text{topo}} = -(\tilde{M}^{AB} + \tilde{M}^{BC} - \tilde{M}^{B} - \tilde{M}^{ABC}),
\label{topomagic}
\end{align}
where $\tilde{M}^{\text{K}}$ denotes the mixed-state stabilizer Rényi entropy (with $n=2$) of the region or subsystem $\text{K} \in \{AB, BC, B, ABC\}$ as shown in Fig.~\ref{fig:intro}(b). 
The superscript $q$ indicates the quadri-partition of the spin chain. This linear combination was first introduced in the context of entanglement as the generalized topological entanglement entropy (TEE) to probe symmetry-breaking orders \cite{zeng2015} as discussed in the supplementary section~\ref{sec:TEE}. 

\subsection{\label{sec:anal}Analytical results}
In the following, we inspect the role of the TSRE in spin chains, starting from analytical solutions of fixed point wave functions. 

\paragraph{Transverse Field Ising chain}
The Hamiltonian of the transverse field Ising model (TFIM) is given by
\begin{align}
\mathcal{H}_I=-J\sum_{\ell=1}^{L-1} Z_\ell Z_{\ell+1}-h\displaystyle\sum_{\ell=1}^{L}X_\ell,
\label{TFI_ham}
\end{align}
where $J$ is the coupling strength between neighboring spins, and $h$ is the transverse magnetic field applied along the $x$-direction. The model possesses a global $\mathbb{Z}_2$ symmetry generated by the operator $\prod_{i=1}^{L}X_i$, corresponding to a collective spin flip in the $x$-basis. As the field strength $h$ is tuned, the system undergoes a quantum phase transition ($h_c=1$) between a ferromagnetic-ordered (FM) phase (for $h<J$) and a paramagnetic-disordered (PM) phase (for $h>J$). This transition is $\mathbb{Z}_2$-protected and considered as symmetry-protected trivial, as indicated by the topological entanglement entropies study in the Appendix~\ref{sec:TEE}. It is therefore natural to investigate the behavior of $M^{q}_{\rm topo}$ in the ground states of $H_I$.

At $h=0$, the ground state of the TFIM within the ordered sector is the Greenberger–Horne–Zeilinger (GHZ) state:
\begin{equation}
|GHZ\rangle = \frac{1}{\sqrt{2}}\left(|00\cdots0\rangle + |11\cdots1\rangle\right)    
\end{equation} for any finite system.
This is a stabilizer state, fully specified by $L$ independent stabilizer generators. For instance, for $L=6$, a valid set is $\{Z_1Z_6,\;Z_2Z_6,\;Z_3Z_6,\;Z_4Z_6,\;Z_5Z_6,\;X_1X_2\dots X_6\}$ which generate a stabilizer group of $2^L$ elements uniquely defining the GHZ state. Being a stabilizer state, its stabilizer Rényi entropy (SRE) vanishes: $M_2 = 0$, since each stabilizer contributes $\pm 1$(Eq.~\ref{eq:sre1}).
Applying a phase gate $S(\theta)=|0\rangle\langle0| + e^{i\theta}|1\rangle\langle1|$ to each qubit produces a doped GHZ state ($|GHZ_\theta\rangle=\frac{1}{\sqrt{2}}\left(|000\cdots0\rangle + e^{iL\theta}|111\cdots1\rangle\right)$. This may alter the state’s magic, while leaving the entanglement structure unchanged. Due to the periodicity of $e^{iL\theta}$, the doped state can remain a stabilizer state for certain $\theta$. The SRE of this doped state is given by
 \begin{align}
M_2(|GHZ_\theta\rangle)=-\log_2\Big(\frac{1+cos^4L\theta+sin^4L\theta}{2}\Big).
\end{align}
In particular, for $\theta = \pi/4$ (i.e., when the phase gate becomes the T-gate), the behavior depends on whether the number of gates is even or odd. 

For even $L$, although the T-gate modifies the phase of the $|11\cdots1\rangle$ component, the stabilizer structure can still be preserved. This is because the T-gate commutes with $Z$, i.e., $TZT^{\dagger}=Z$, leaving generators of the form $Z_i Z_L$ unchanged. The global $X^{\otimes L}$ generator, however, is no longer a single stabilizer; its conjugation under the T-gate produces a linear combination of $2^L$ Pauli strings (as $T X T^{\dagger} = \frac{X+Y}{\sqrt{2}}$), which still generate a valid stabilizer group and can reconstruct the original $Z_i Z_L$–type stabilizers. A possible generating set in this case (for $L=6$) is $\{Y_1X_2\cdots X_6,\;X_1Y_2X_3\cdots X_6,\;X_1X_2Y_3X_4X_5X_6$, $X_1X_2X_3Y_4X_5X_6,\; X_1\cdots X_4Y_5X_6,\;X_1\dots X_5Y_6\}$. These generators form a complete and independent stabilizer group, indicating that the doped GHZ state remains a stabilizer state, albeit in a rotated/different stabilizer basis. As a result, the SRE again vanishes ($M_2=0$). 

In contrast, for odd $L$, the stabilizer group is reduced to $2^{(L-1)}$ elements contributing $\pm1$ since $Z_iZ_L$ generators still commute with T-gate. However, the global generator ($X^{\otimes L}$) transform into linear combination of $2^L$ Pauli strings, which contribute $\pm (1/\sqrt{2})$ to the SRE trace. Unlike the even $L$ case, these stabilizers do not close under multiplication to form a stabilizer group, which indicates that the state is no longer a pure stabilizer state, but rather a superposition of stabilizer states, with SRE: $M_2(|GHZ_T\rangle)=\log_2(4/3)\approx 0.415$.

Importantly, the GHZ states remain locally indistinguishable from stabilizer states for any subsystem. This follows from their high degree of symmetry under global bit flips and their purely classical correlation structure in the computational basis. When restricted to any subsystem, the stabilizer group of the GHZ state contains only identity and  $Z$-type operators, which contribute trivially to the local Pauli correlators entering in the equation of mixed state SRE defined in Eq.~\ref{Mix_SRE}.
As a result, these contributions are exactly canceled by the normalization term in the denominator, yielding a vanishing mixed-state stabilizer Rényi entropy,
$\tilde{M}_2(\rho_A(\theta))=0$. Consequently, both the undoped and the $T$-doped GHZ states remain locally stabilizer-like, exhibiting no subsystem magic.
For the quadripartition geometry (Fig.~\ref{fig:intro}(b)), all partitions contribute zero magic, giving $M^q_{\rm topo} = 0$ from Eq.~\eqref{topomagic}. This confirms that the GHZ state is topologically trivial, with no nontrivial entanglement structure or TSRE.

In the opposite limit $h\to\infty$, the ground state of the model is a product state in the $x$-basis: $|+\rangle^{\otimes L}$, with $|+\rangle=\frac{1}{\sqrt{2}}(|0\rangle+|1\rangle)$,  which is a stabilizer state, fully characterized by $L$ independent generators. 
For example, when $L=8$, the generators are: $\{X_1,\;X_2,\;\dots,\;X_7,\;X_8\}$. These generate a stabilizer group with $2^L$ elements, and hence the state has zero magic (Eq.~\ref{eq:sre1}). When phase gates are applied to each qubit, the product state transforms into a doped product state: $|+_\theta\rangle^{\otimes L}=\prod_iS_i(\theta)|+\rangle^{\otimes L}$ which is no longer a stabilizer state. This happens because the phase (or T) gates do not commute with the Pauli $X$ operators as $TXT^\dagger=\frac{1}{\sqrt{2}}(X+Y)$. As a result, each $X_i$ generator is transformed to a superposition of Pauli strings involving both $X$ and $Y$. This generates a total of $3^L(=1+\sum_k C_{k}^L 2^k)$ distinct Pauli strings, which do not form a stabilizer group and hence the resulting state is no longer a stabilizer state. The total contribution of these Pauli strings to the SRE is given by
\begin{align}
\sum_{P\in\mathcal{P}_L}\hspace{-4pt}|\langle\psi|P|\psi\rangle|^4=\hspace{-8pt}\sum_{\substack{n_I+n_X\\+n_Y=L }}\hspace{-5pt}\frac{L!}{n_I!n_X!n_Y!}1^{n_I}(\cos^{4}\theta)^{n_X}(\sin^{4}\theta)^{n_Y},
\end{align}
where $n_I$, $n_X$, $n_Y$ count the occurrences of $I,X,Y$ in a given Pauli string, respectively. The resulting magic for the doped state is 
\begin{align}
M_2(|\psi_\theta\rangle)=-L\log_2\Big( \frac{1 + \cos^4\theta + \sin^4\theta}{2} \Big).
\end{align}
For a subsystem $K$ of length $L_K$, the mixed-state SRE becomes:
\begin{align}
\tilde{M}_2(|\psi_\theta\rangle) = -L_K \log_2\Bigg( \frac{1 + \cos^4\theta + \sin^4\theta}{1 + \cos^2\theta + \sin^2\theta} \Bigg),
\end{align}
showing that magic is locally distributed and additive, so the SRE scales linearly with the partition size. In the limiting case of the T-gate ($S(\pi/4)$), it reduces to
$\tilde{M}_2(|\psi_T\rangle)=L_K \log_2(4/3)$. This additivity immediately implies a vanishing TSRE. Since each partition contributes in proportion to its length, the combination appearing in Eq.~\eqref{topomagic} cancels exactly, yielding $M^{q}_{topo}=(L_{ABC}+L_{B}-L_{AB}-L_{BC})\log_2(4/3)=0$. Therefore, like the GHZ state, the paramagnetic state is topologically trivial when characterized by topological magic.

\paragraph{Cluster Ising chain}
The Hamiltonian for the Cluster Transverse Field Ising (CTFI) model~\cite{PhysRevA.84.022304,PhysRevLett.93.056402,son2011quantum} is given by
\begin{align}
\mathcal{H}_{CI}=J\displaystyle\sum_{\ell=1}^{L-2} Z_\ell X_{\ell+1}Z_{\ell+2}+h\displaystyle\sum_{\ell=1}^{L}X_\ell, 
\label{CTFI_ham}
\end{align}
where $J=1$ is the interaction strength, and $h$ is the transverse field applied in the $x$-direction. 
The first term represents three-body cluster-type interactions, while the second corresponds to a uniform magnetic field. 
This model exhibits a $\mathbb{Z}_2\times \mathbb{Z}_2$ symmetry, generated by the operators $X_o=\prod_{i=1}^{L/2}X_{2i-1}$ and $X_e=\prod_{i=1}^{L/2}X_{2i}$. 
The system undergoes a phase transition from the symmetry-protected topological (SPT) phase to a paramagnetic phase, driven by the competition between the three-body term and the transverse field strength. A clear signature of this is captured by the quadri-parition entanglement entropy ($S^{q}_{topo}$) in Fig.~\ref{fig:topoEE} (d)(see Supplementary Sec.~\ref{sec:TEE}).

For $h=0$, the ground state of the system corresponds to the Cluster (CL) state, which has been widely studied as a resource for quantum computation~\cite{PhysRevLett.86.910,NIELSEN2006147}. The ground state is four-fold degenerate, with the degenerate states connected by the global symmetry operations $X_o$, $X_e$, and $X_oX_e$, where $X_o$ and $X_e$ are the products of Pauli $X$ operators on odd and even sites, respectively. The symmetric superposition of these degenerate states defines the full ground state:
\begin{align}
|\text{CL} \rangle = 
\begin{cases} 
\frac{(I+X_e+X_o+X_eX_o)}{2}|\psi_0\rangle & \text{for even } L, \\
\frac{(I + X_oX_e)}{\sqrt{2}}|\psi_0\rangle & \text{for odd } L.
\end{cases}
\end{align}
Here the reference state ($|\psi_0\rangle$)~\cite{skeleton2021,PhysRevLett.97.110403} is constructed using the MPS structure given by 
\begin{align}
|\psi_0\rangle=\frac{1}{2^{L/2}}
\begin{pmatrix}
1 &0
\end{pmatrix}
\begin{pmatrix}
|1\rangle &-|1\rangle\\
|0\rangle &|0\rangle
\end{pmatrix}^{L}\begin{pmatrix}
1\\-1
\end{pmatrix}.
\end{align}
The final Cluster state is a pure stabilizer state, fully characterized by a group of $2^L$ Pauli operators generated from a set of $L$ independent stabilizer generators. For instance, for a chain of length $L=8$, one possible choice of stabilizer generators is $\{Z_1Z_7X_8,\;Z_2X_3X_5X_7Z_8,\;Z_3X_4X_6Z_7,\;Z_4X_5X_7Z_8$, $Z_5X_6Z_7,\;Z_6X_7Z_8,\;X_1X_3X_5X_7,\;X_2X_4X_6X_8\}$ and for $L=7$ is $\{Z_1Z_7,\;Z_2X_3X_5Z_6,\;Z_3X_4X_6Z_7$, $Z_4X_5Z_6,\;Z_5X_6Z_7,\;X_1X_3X_5X_7,\;X_2X_4X_6\}$.
Since the entire state lies within the stabilizer formalism, its stabilizer Rényi entropy (or magic) is exactly zero. 

However, when a $T$-gate is applied to each qubit ($N_T=L$), the state is transformed into a doped cluster state that generally lies outside the stabilizer formalism, as original stabilizer generators do not commute with the $T$-gate. For even $L$, none of the original generators commute with the $T$-gate, and the number of resulting Pauli strings after superposition is  $3^L$. Instead, for odd $L$, there exists a $Z_1Z_L$-type generator that commutes with all $T$-gates, reducing the number of contributing strings to $2^23^{L-2}$. 
Thus, similar to the GHZ state, the SRE of the doped cluster state depends on whether the system size is even or odd. The full state SRE is given by 
\begin{align}
M_2(|\text{CL}_{T} \rangle) = 
\begin{cases} 
L \log_2(4/3) & \text{for even } L, \\
(L-2) \log_2(4/3) & \text{for odd } L.
\end{cases}
\end{align}
The stabilizer Rényi entropy (SRE) of a subsystem depends not only on its size but also on the connectivity of different partitions of it. Similar to entanglement entropy, connected and disconnected regions behave differently: for connected partitions, the SRE scales as $L_K - 2$, while for disconnected ones it scales as $L_K - 4$. This originates from the symmetry and correlation structure of the Cluster state (with or without T-gates), preserved under the global operators ($X_e, X_o, X_eX_o$). These stabilizers encode nonlocal correlations between even and odd sites, which are disrupted once by a connected cut and twice by a disconnected cut. Consequently, the doped Cluster state exhibits non-trivial topological SRE: 
\begin{align}
M^q_{topo}=[-(L_{AB}-2)-(L_{BC}-4)+(L_B-2)\nonumber\\+(L_{ABC}-2)]\log_2(\frac{4}{3})=2\log_2(\frac{4}{3}).
\label{eq:CluSRE}
\end{align}
Note that in the quadri-partition, the $ABC$ region is formally disconnected but still contributes as if connected because the topological edge modes remain nonlocally linked. Oppositely, the $BC$ region is truly disconnected and breaks these links. 

The behavior of the TSRE closely parallels that of topological entanglement entropy in quadripartitions of the Cluster state, where a universal contribution of $2\log_2 2$ appears.
By contrast, the undoped ground state exhibits vanishing quadripartition magic, consistent with its symmetry-protected stabilizer character. This unequivocally illustrates how magic response is required to establish a connection between non-stabilizerness and topology.

In the limit $h\to\infty$, the system enters a paramagnetic phase with vanishing magic, analogous to the large-field regime of the transverse field Ising model, as discussed previously. Doping with T-gates thus distinguishes the two phases and provides better topological insights from the perspective of magic. A summary of these analytical findings for both models is provided in Table~\ref{table}.

\begin{table}[h!]
\centering
\renewcommand{\arraystretch}{1.4} 
\begin{tabular}{|c|c|c|c|}
\hline
\rowcolor{gray!40}
\multicolumn{1}{|c|}{\parbox{4cm}{\centering \diagbox[height=2em]{\;\;\textbf{Quantity}}{\textbf{States}\;\;}}} 
&\;\;\textbf{PM}\;\; & \;\;\textbf{FM}\;\; & \;\;\textbf{CL}\;\; \\
\hline
$S^q_{\text{topo}}$ & 0 & 0 & $2\log_2(2)$ \\
\hline
$M^q_{\text{topo}},\ N_T=0$ & 0 & 0 & 0 \\
\hline
$M^q_{\text{topo}},\ N_T=1$ & 0 & 0 & 0 \\
\hline
$M^q_{\text{topo}},\ N_T=L$ & 0 & 0 & $2\log_2(4/3)$ \\
\hline
\end{tabular}
\caption{Similarity of topological quantities in quadri-partition setup: entanglement entropy ($S^{q}_{\text{topo}}$) and topological stabilizer R\'enyi entropy ($M^{q}_{\text{topo}}$) for paramagnetic (PM), ferromagnetic (FM), and cluster (CL) phases in symmetry-protected trivial and topological phases, with and without T-gates, where $N_T$ is the number of T-gates and $L$ is the system size.}
\label{table}
\end{table}

\begin{figure}[t]
    \centering
    \includegraphics[width=0.9\linewidth]{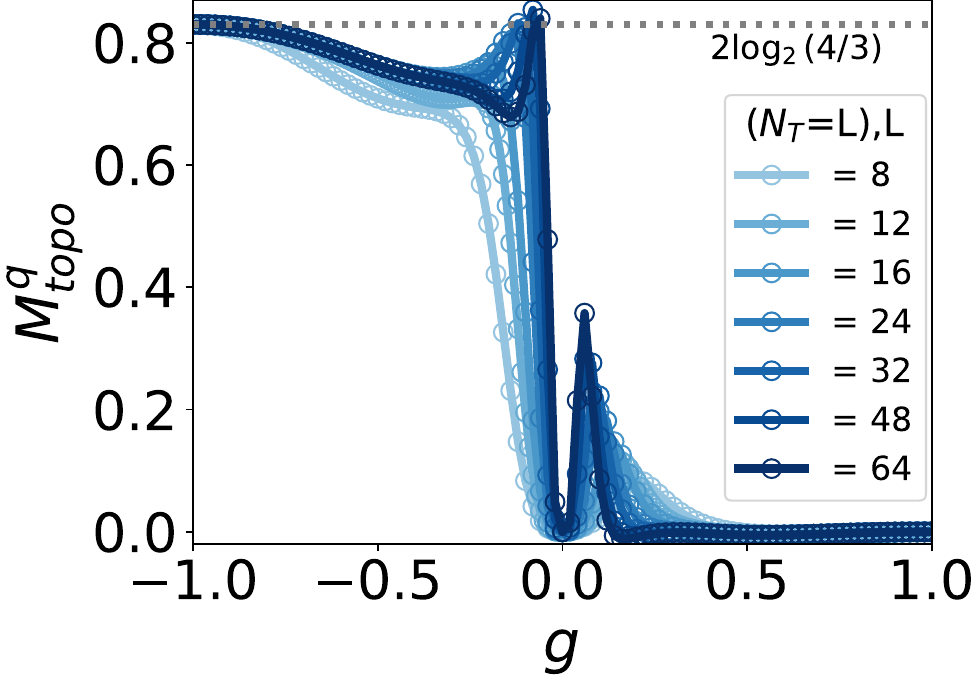}
\caption{The quadri-partition topological stabilizer Rényi entropy (TSRE), $M^q_\mathrm{topo}$, as a function of $g$ in the tri-critical Ising model (Eq.~\ref{TCIM_ham}). The results are computed using the MPS representation of the fully T-gate doped state ($N_T=L$), $|\psi_T\rangle$ (Eq.~\ref{MPS_Tgate}), for different system sizes $L$.}
\label{fig:TCIsing}
\end{figure}

\begin{figure*}
\centering
\includegraphics[width=0.323\linewidth]{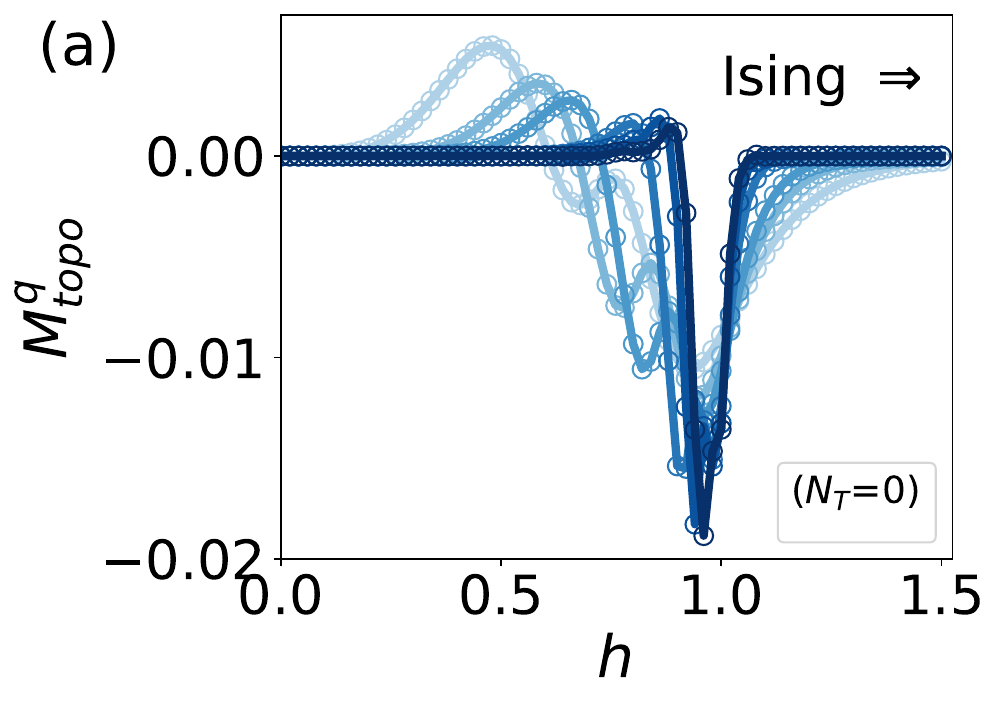}
\includegraphics[width=0.323\linewidth]{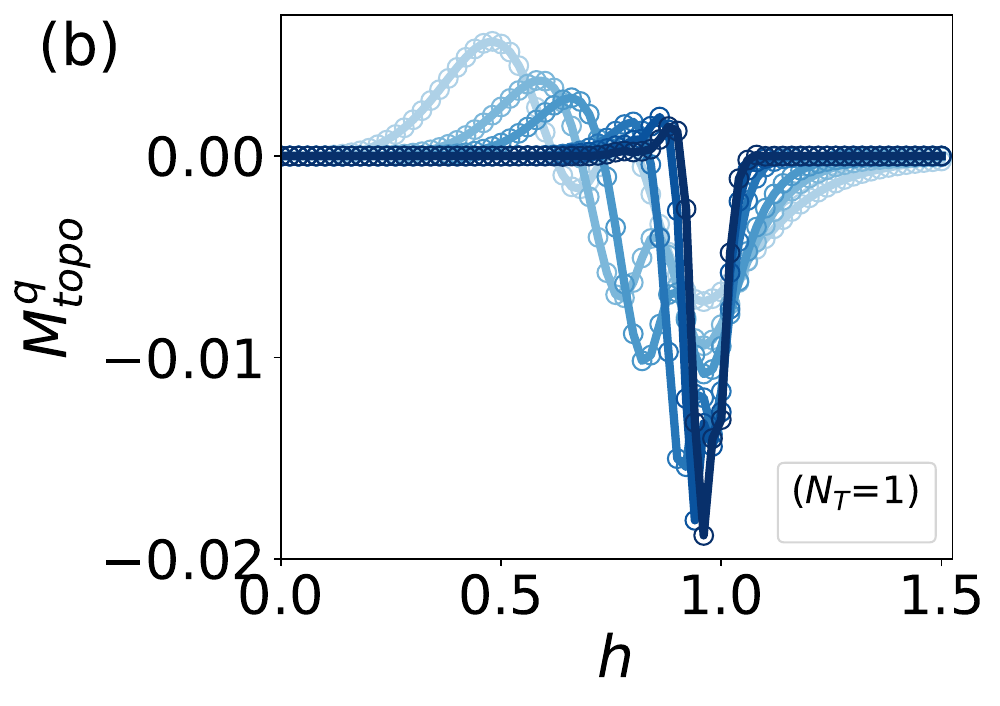}
\includegraphics[width=0.323\linewidth]{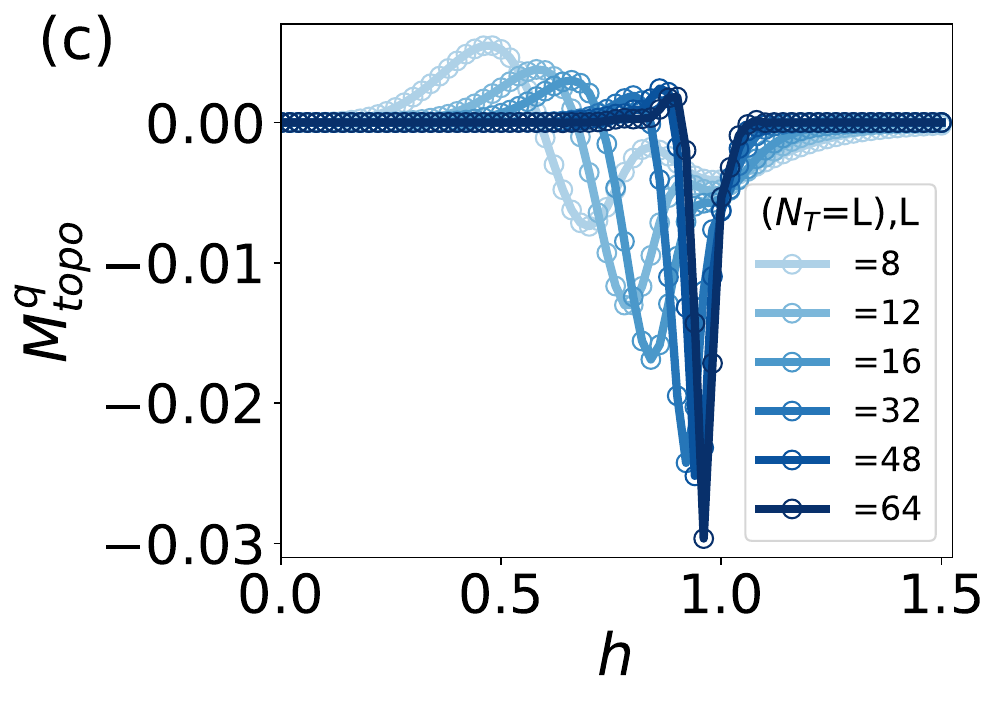}\\
\includegraphics[width=0.323\linewidth]{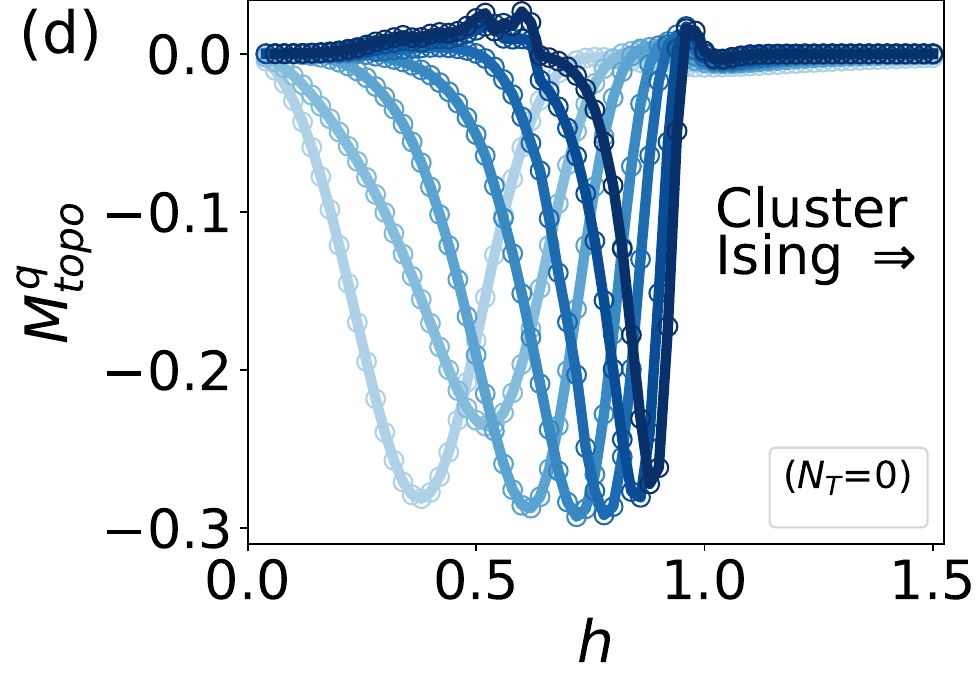}
\includegraphics[width=0.323\linewidth]{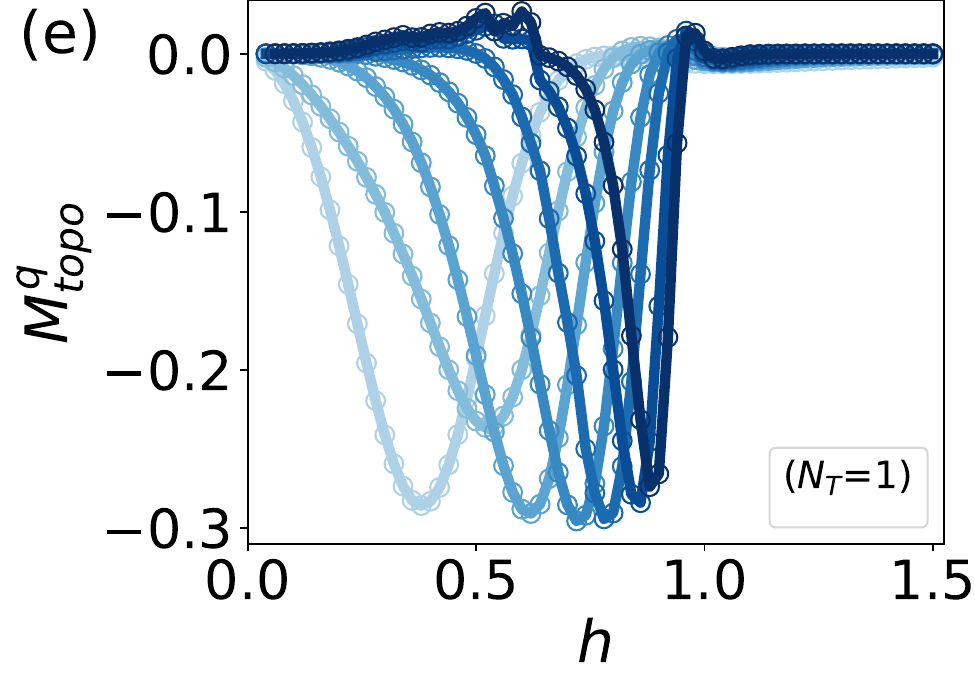}
\includegraphics[width=0.323\linewidth]{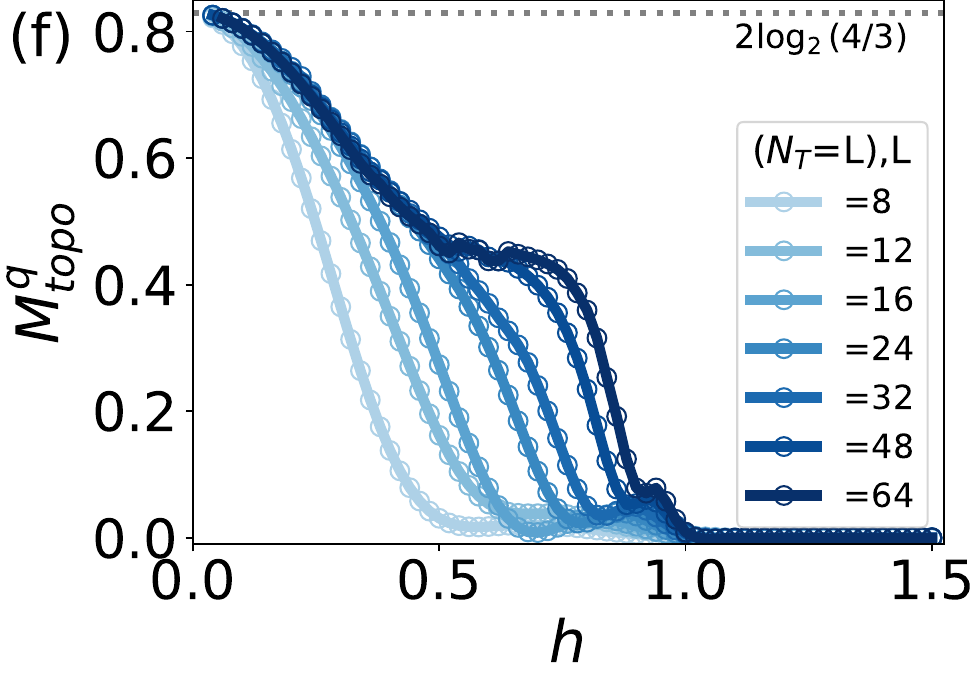}
\caption{Topological stabilizer Renyi entropy ($M^q_{topo}$) of the (a,b,c) Ising model and (d,e,f) Cluster Ising model with (a,d) $N_T=0$, (b,e) $N_T=1$ and (c,f) $N_T=L$ number of T-gates on the ground state of system size $L$ as a function of transverse field $h$ with $J=1$.}
\label{fig:quad_Magic}
\end{figure*}

\subsubsection{\label{sec:TCI}A fully solvable phase diagram: Tri-critical Ising model}

While the analysis above just focuses on fixed points, it is important to have an semi-analytical understanding of how topological SRE scales over the full phase diagram of an interacting spin chain. For this purpose, we investigate the ground-state properties of the tri-critical Ising (TCI) model. The three-term Hamiltonian of the TCI model can be expressed in terms of a single parameter $g$ as~\cite{skeleton2021,PhysRevLett.97.110403,PhysRevResearch.4.L022020}
\begin{align}
\mathcal{H}_{\text{TCI}} = 2(g^2-1)\sum_{\ell=1}^{L-1} Z_\ell Z_{\ell+1}
-(g+1)^2\sum_{\ell=1}^{L}X_\ell \nonumber \\
+(g-1)^2\sum_{\ell=1}^{L-2} Z_\ell X_{\ell+1}Z_{\ell+2},
\label{TCIM_ham}
\end{align}
where $\{X, Z\}$ are Pauli operators. The model preserves both a global spin-flip symmetry, generated by $\prod_{i=1}^{L}X_i$, and time-reversal symmetry. As a result, it hosts three distinct phases: a $\mathbb{Z}_2 \times \mathbb{Z}^T_2$ symmetry-protected topological (SPT) phase, a trivial phase, and a symmetry-broken phase. At special values of $g$, the ground state reduces to a Cluster state for $g=-1$, to a paramagnetic state for $g=1$, and at the critical point $g=0$ it corresponds to the GHZ state.

The matrix product state (MPS) representation of the wavefunction at each site can be written using two matrices,
\begin{align}
A_0=\begin{pmatrix}
0 &0\\
1 &1
\end{pmatrix},\;A_1=
\begin{pmatrix}
1 &g\\
0 &0
\end{pmatrix}\equiv A =
\begin{pmatrix}
|1\rangle & g|1\rangle\\
|0\rangle &|0\rangle
\end{pmatrix}.
\end{align}
For a chain of even length $L$ under open boundary conditions, the ground state can be expressed as a symmetric superposition of two normalized MPS wavefunctions given as 
\begin{align}
|\psi\rangle=\frac{1}{\sqrt{2N_m}}\text{tr}(|\mathcal{L}\rangle \langle \mathcal{R}| A^L),N_m=\frac{(1+g)^L+(1-g)^L}{2},
\end{align}
where $|\mathcal{L}\rangle$ and $\langle \mathcal{R}|$ are left and right boundary vectors such that $|\mathcal{L}\rangle \langle \mathcal{R}|=Z$.
After doping the entire chain with T-gates, the MPS representation of the state becomes
\begin{align}
|\psi_T\rangle=\frac{1}{\sqrt{2N_m}}\text{tr}(|\mathcal{L}\rangle \langle \mathcal{R}|A_T^L); \quad  A_T =
\begin{pmatrix}
e^{i\frac{\pi}{4}}|1\rangle & ge^{i\frac{\pi}{4}}|1\rangle\\
|0\rangle &|0\rangle
\end{pmatrix}.
\label{MPS_Tgate}
\end{align}
Using this MPS representation of the doped ground state $|\psi_T\rangle$, we compute the exact topological SRE for quadri-partition, $M^q_{\rm topo}$ of the Tri-critical Ising model, as shown in Fig.~\ref{fig:TCIsing}.

As in the standard and Cluster Ising models, the tri-critical Ising model exhibits non-trivial topological features within the topological phase, while the topological SRE vanishes in the trivial phase. In the thermodynamic limit, the topological SRE is expected to reach the constant value $2\log_2(4/3)$, as the MPS representation becomes exact and free of truncation errors. For special values of $g$, the ground state simplifies to known states. For $g=-1$, it corresponds to an MPS representation of the Cluster state and shows $M^q_{\rm topo}=2\log_2(4/3)$, as shown in Fig.~\ref{fig:TCIsing};  at $g=0$ and $g=1$, corresponding to the GHZ and paramagnetic states, respectively, for which topological SRE ($M^q_{topo}$) vanish, also illustrated in Fig.~\ref{fig:TCIsing}. These exact results are fully consistent with the analytical findings for the Ising and Cluster Ising models discussed in the previous subsections, and they provide a precise characterization of the non-trivial behavior in the symmetry-protected topological (SPT) phases due to the exact, truncation-free MPS representation.

\subsection{Numerical Results}
After the analytical study, we present the numerical results for the quadripartition topological SRE ($M^q_{\rm topo}$), obtained using the Pauli MPS method (see Methods section~\ref{sec:method}). The subsystem partitions are defined as
$A=[1,L/4]$, $B=(L/4,L/2]$, $D=(L/2,3L/4]$, and $C=(3L/4,L]$, as illustrated in Fig.~\ref{fig:intro}(b).

\subsubsection{Spin-1/2 chain}

We start our analysis by considering topological SRE without magic response. The corresponding results are depicted in Fig.~\ref{fig:quad_Magic}(a,d), for the Ising and Cluster Ising model, respectively. In both cases, topological SRE vanishes as volume is increased, with differences only at the quantitative level. These results are fully in line with expectations from the fixed point computations of the previous section: even in the SPT phase of the cluster Ising chain, the fixed point description is a stabilizer state. Similar results are obtained in the presence of a single T-gate (Fig.~\ref{fig:quad_Magic}(b,e)).

We then investigate topological magic response, by doping the ground states with $N_T=L$ number of T-gates. The corresponding results are shown in Fig.~\ref{fig:quad_Magic}(c,f). This doping enhances the magic across all parts of the chain, making the two SRE contributions ($\tilde{M}^{AB}, \tilde{M}^{BC}$) clearly distinguishable throughout the SPT phase of the cluster Ising chain, while they again remain identical in the trivial regimes. Consequently, $M^q_{\rm topo}$ remains zero in both paramagnetic and ferromagnetic phases (Fig.~\ref{fig:quad_Magic}(c,f)), but becomes nonzero in the Cluster phase. This behavior is fully consistent with the analytical studies of these states discussed earlier.

Importantly, the doped Cluster state exhibits a topological SRE of order $2\log_2(4/3)$, corresponding to the contribution of two T-gates, closely paralleling the universal value $2\log_2(2)$ of the topological entanglement entropy in the SPT phase (Fig.~\ref{fig:topoEE}(c,d)). This value, however, does not remain constant across the phase: in fact, our numerics are consistent with a finite size scaling to a value that is always smaller than $2\log_2(4/3)$ in most of the phase, as exemplified by a scaling presented in Fig.~\ref{fig:intro}(a). As the system approaches the critical point, local correlations generate additional types of magic, thereby reducing the T-gate contribution. 

\subsubsection{Resilience to disorder} 

To further assess the robustness of topological SRE, we examine how the latter responds to disorder in the cluster Ising model. We introduce random nearest-neighbor couplings into the Hamiltonian of Eq.~\eqref{CTFI_ham}, obtaining
\begin{align}
\mathcal{H}_{CID}=J\displaystyle\sum_{\ell=1}^{L-2} Z_\ell X_{\ell+1}Z_{\ell+2}+h\displaystyle\sum_{\ell=1}^{L}X_\ell+ \sum_{\ell=1}^{L-1}\Delta_{\ell}Z_\ell Z_{\ell+1},
\label{CTFI_dis_ham}
\end{align}
where each $\Delta_\ell$ is independently sampled from a uniform distribution $\Delta_\ell \in [-\Delta, \Delta]$. The type of the disorder preserves the underlying $Z_2 \times Z_2$ symmetry and thus maintains the topological character of the cluster phase for small disorder amplitudes.
\begin{figure}[t]
\centering
\includegraphics[width=0.8\linewidth]{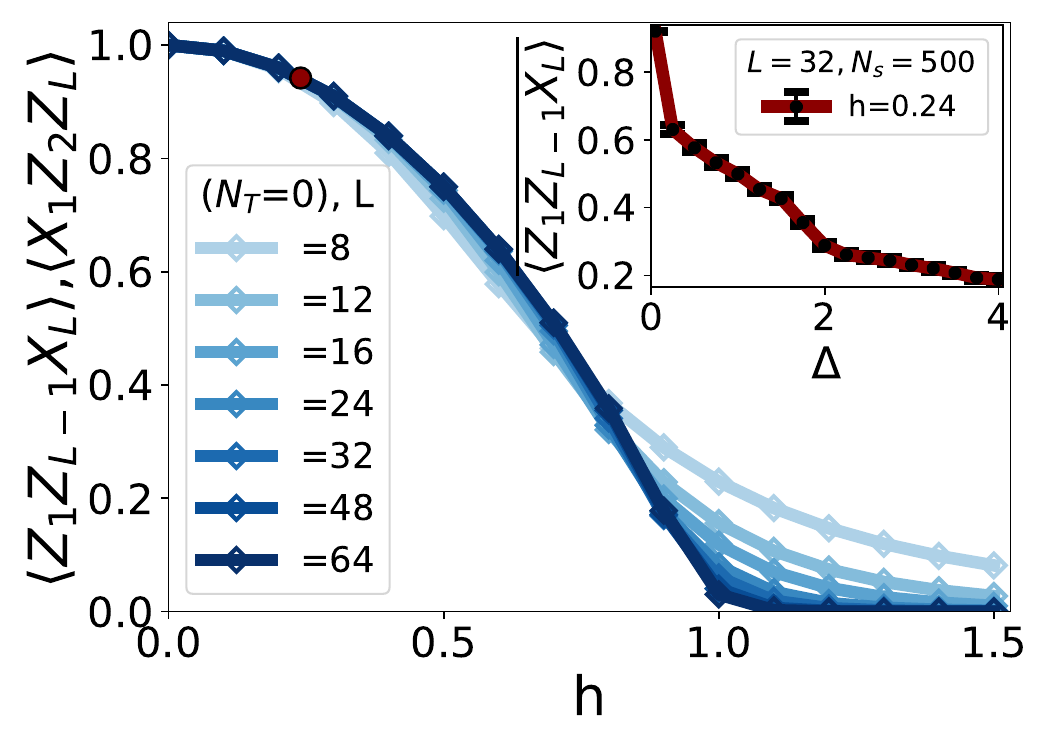} 
\caption{The two edge modes correlation as a function of the magnetic field strength ($h$) in the Cluster Ising model for different system sizes $L$, in the absence of T-gates and $J=1$. The inset shows the effect of disorder strength ($\Delta$) on the correlation of a single edge mode for $L=32$ and $h=0.24$ (corresponding to the red dot in the main figure), averaged over $N_s=500$ disorder realizations.}
\label{fig:edge_corl}
\end{figure}

In the clean limit, the topological cluster phase is signaled by long-range correlations between the edge modes, which decay only upon approaching the phase transition to the trivial regime at large field $h$ (Fig.~\ref{fig:edge_corl}).
These edge correlations are directly mirrored in the topological SRE, ($M^q_{\rm topo}$), which remains finite throughout the topological phase even when the ground state is locally perturbed by T-gates (Fig.~\ref{fig:quad_Magic}(f)). Upon introducing disorder, additional short-range correlations are generated, progressively diminishing the non-local edge correlations and ultimately eroding topological protection. This crossover is illustrated in the inset of Fig.~\ref{fig:edge_corl} for a representative set of parameters.

\begin{figure}[t]
\centering
\includegraphics[width=0.8\linewidth]{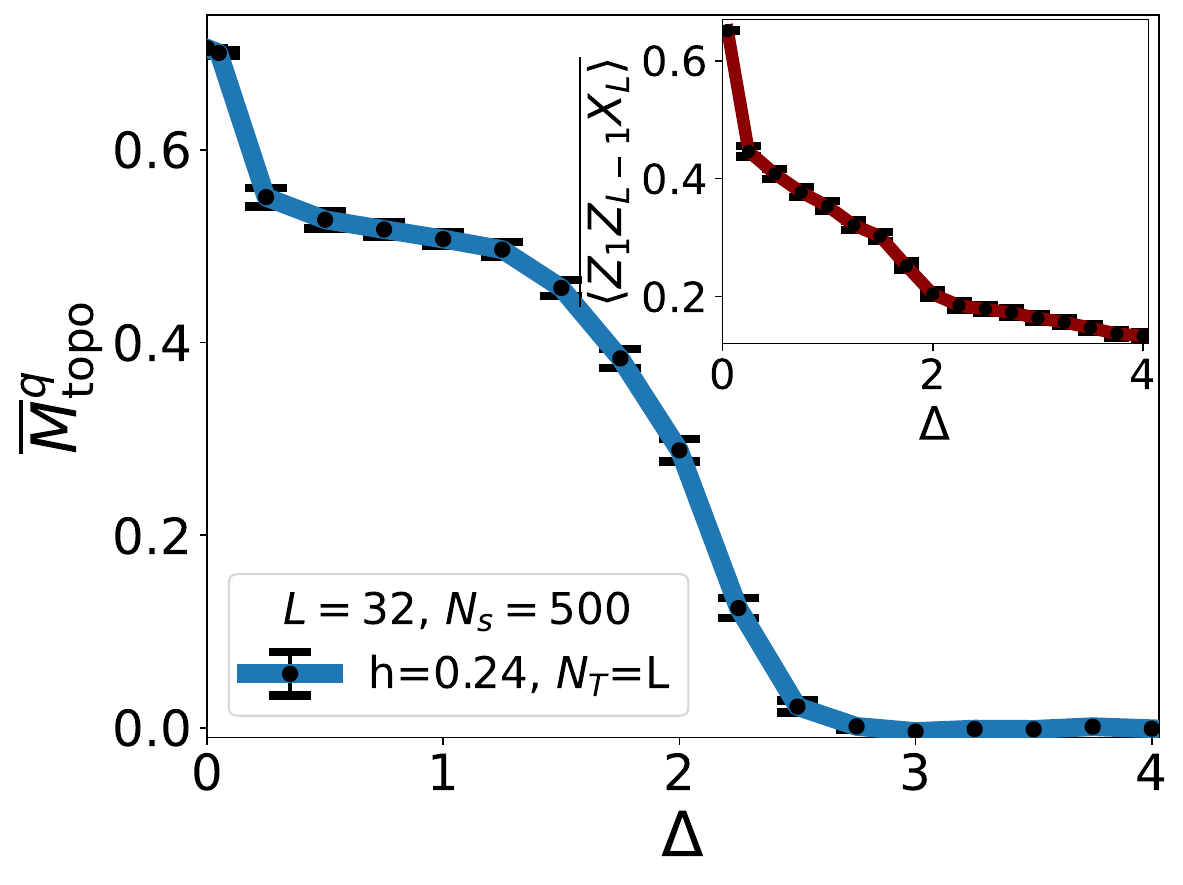} 
\caption{Topological SRE $M^q_{\rm topo}$ as a function of disorder strength $\Delta$ for the $N_T = L$ T-gates doped ground state of the Cluster Ising model, with $h=0.24$, $J=1$, and $L=32$. The inset shows the correlation of a single edge mode under the same conditions, averaged over $N_s=500$ disorder realizations, highlighting the effect of increasing disorder.}
\label{fig:SRE_dis}
\end{figure}

The topological SRE captures the same behavior as shown in Fig.~\ref{fig:SRE_dis}: $M^q_{topo}$ is remarkably stable against weak disorder, but gradually suppressed once the disorder exceeds a critical strength, in close correspondence with the decay of the edge–edge correlations (inset of Fig.~\ref{fig:SRE_dis}). These results demonstrate that $M^q_{topo}$ provides a reliable and disorder-resilient probe of the topological phase — remaining robust to local perturbations as long as the underlying symmetry is preserved, and decaying only when disorder becomes strong enough to destroy the non-local structure of the SPT state.

\subsubsection{Spin-1 chain}
Alongside the SPT phases in spin-1/2 systems, we study topological SRE in the spin-1 setting. To investigate the topological properties of the spin-1 AKLT state from the perspective of magic, we consider a one-dimensional spin-1 Hamiltonian discussed in Ref.~\cite{PhysRevLett.121.140604} whose ground state interpolates between the Affleck-Kennedy-Lieb-Tasaki (AKLT) state and a trivial state $|0\rangle^{\otimes N}$,
\begin{align}
H_{AKLT}= (1-\delta)\displaystyle\sum_{\ell=1}^{L-1} \left[\mathbf{S}_\ell \cdot \mathbf{S}_{\ell+1}
+ \frac{1}{3}\left(\mathbf{S}_\ell \cdot \mathbf{S}_{\ell+1}\right)^2 \right]\nonumber \\
+ \delta \sum_{\ell=1}^{L} (S^z_\ell)^2.
\label{eq:hamAKLT}
\end{align}
Here $\mathbf{S}_\ell = (S^x_\ell, S^y_\ell, S^z_\ell)$ are spin-1 operators on on site $\ell$. 
The parameter $\delta$ controls the relative strength between the AKLT Hamiltonian and the single-ion anisotropy. As $\delta$ increases, the system interpolates smoothly between the Haldane phase at $\delta=0$, and a trivial, large-D phase for sufficiently large $\delta$. Around $\delta \approx 2/3$, the model undergoes a continuous quantum phase transition separating these two regimes. This phase transition is protected by the underlying $U(1)$ symmetry generated by $\sum_j S^z_j$ as well as by time-reversal and inversion symmetries, which together ensure the stability of the Haldane SPT phase against symmetric perturbations. In the topological regime $\delta \lesssim 2/3$, the system hosts fractionalized edge excitations and non-local string order, whereas for $\delta \gtrsim 2/3 $,  these features disappear and the ground state becomes adiabatically connected to a product state. This makes the model an ideal testbed for investigating how topological SRE behaves across an interacting symmetry-protected phase transition. 
\begin{figure}[t]
    \centering
    \includegraphics[width=0.9\linewidth]{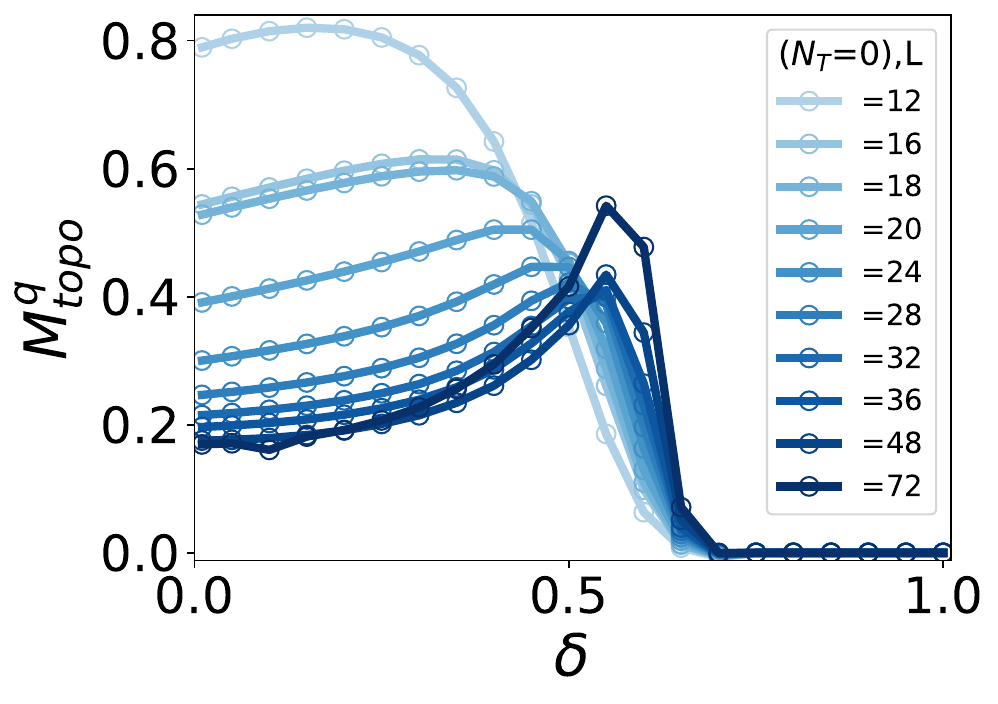}
    \caption{The quadri partition topological Stabilizer R\'{e}nyi entropy of the ground state of the spin-1 AKLT model as a function of tuning parameters $\delta$. }
    \label{fig:SREAKLTq}
\end{figure}
This phase transition is nicely captured by the study of topological entanglement entropy ($S^q_{topo}$) as shown in Fig.~\ref{fig:topoEE}(f) in the Appendix \ref{sec:TEE}.

Spin-1 systems inherently possess more quantum magic than spin-1/2 models, and thus non-trivial features can emerge even in the absence of non-Clifford T-gates. Consequently, the quadri-partition SRE of this model (Eq.~\ref{eq:hamAKLT}) reflects this intrinsic non-triviality, as shown in Fig.~\ref{fig:SREAKLTq}, even without any additional T-gate doping of the ground state. 
All numerical results were obtained from ground-state MPS simulations with a truncation cutoff of $10^{-15}$ and a maximum bond dimension of $\chi_P=55$. The cutoff used for computing the various SREs is $10^{-12}$.

A closer examination of Fig.~\ref{fig:SREAKLTq} reveals that the SRE vanishes completely in the trivial region, while in the topological phase it approaches a constant value ($\approx 0.169$) in the thermodynamic limit. This behavior arises primarily from the reduced magic in the partition $BC$  compared to $AB$, despite both having the same length. The difference stems from their distinct topological roles; the $BC$ partition includes the symmetry-protected edge contribution, which suppresses its magic relative to $AB$, consistent with what is also observed in spin-1/2 models. We observe that, for any partition, the quantum magic remains largely constant throughout the Haldane phase, with the plateau value depending on the partition size. After crossing the transition point, however, the magic decreases and eventually vanishes in the trivial region as the quantum state approaches the product state ($|0\rangle^{\otimes L}$), ultimately becoming a stabilizer state.

This behavior contrasts sharply with that of spin-1/2 models, where the ground states begin as stabilizer states and the magic slowly increases, finally reaching its maximum near the transition point. Consequently, the distinction between trivial and non-trivial phases is not as prominently captured in spin-1/2 systems, whereas it emerges clearly in the spin-1 model, even without the addition of T-gates. Since spin-1 states already possess a significant amount of quantum magic, doping the ground states with T-gates has little effect on the topological SRE, producing only a small overall increase in the constant value of magic. One possible explanation for this fact is that, for odd prime Hilbert space dimensions, SREs are in fact directly related to measures of magic~\cite{Leone2022stabilizer,tarabunga2024critical}, and thus potentially more faithfully sensitive to quantum correlations. 

\section{Discussion}

We have shown how non-locally encoded information in topological quantum matter can have a direct connection to magic, introducing the concept of magic response - that is, how much a given state is able to distribute over stabilizer space upon the action of a local rotation that leaves entanglement unaffected. Our analytical and numerical computations establish a sharp  
dichotomy among quantum phases of one dimensional spin chains. Trivial and symmetry-broken phases, such as the ferromagnetic and paramagnetic regimes of the Ising chain, remain stabilizer-like even after uniform T-gate doping: their magic is purely local, and their topological stabilizer R\'{e}nyi entropy vanishes for all system sizes. By contrast, SPT phases—exemplified by the Cluster Ising model and the SPT regime of the tri-critical Ising model—display a pronounced magic response. As a sum up, the main contribution of our work is to point out that the relation between magic and topology is quite distinct from that between the latter and entanglement: it is related to how our states can respond to magic perturbations, rather than how magic is distributed within the state itself.

There are multiple questions our work opens. The first one is, whether it is possible to formulate a topological invariant from magic. The very same question is still open even for entanglement~\cite{zou2016spurious}: in fact, topological entanglement entropies are rigorous lower bounds~\cite{levin2024physical}, and the formulation of entanglement topological invariants is in progress, utilizing concepts beyond entropies~\cite{kim2023universal}. It might be possible to leverage the flexibility of magic response by, e.g., considering a broad set of rotations. On a different route, the stable plateaus we observed for spin-1 systems may signal that a proper measure of magic, such as robustness, could provide a clearer picture for qubits as well. Another interesting extension is to go towards topological orders and, in general, gauge theories on lattices, for which relatively little is known. Understanding how topological stabilizer R\'{e}nyi entropies work in that context might also shed light on the working on Clifford augmented tensor networks, which is related~\cite{10.21468/SciPostPhys.18.5.165,PhysRevB.111.085121} to the presence of non-local transformations similar to those found in lattice gauge theory~\cite{fradkin2013field}.

\paragraph*{Acknowledgements} 
We thank B. Beri and H. Timsina for useful discussions. M. D. acknowledges inspiring feedback from L. Piroli, in particular, in the context of resilience to perturbations. 
E.T. thanks X. Turkeshi, P. Sierant and D. Bhakuni for insightful discussions and collaboration on related projects.
M.~D. acknowledges funding from the European Research
Council (ERC) under the European Union’s Horizon ERC-2022-COG Grant with Project number 101087692,  and the PRIN programme (project CoQuS). 
M.~C.and M.~D. acknowledges funding from
the PNRR MUR project PE0000023-NQSTI.
R. N. and M.~D. were also supported by the EU-Flagship programme Pasquans2.
E.T. acknowledges funding from the Swiss National Science Foundation (SNSF) under Grant No. TMPFP2234754, and
CINECA (Consorzio Interuniversitario per il Calcolo Automatico) award, under the ISCRA initiative and Leonardo early access program, for the availability of high-performance computing resources and support. P.S.T. acknowledges funding from the European Research
Council (ERC) under the European Union (ERC, DynaQuant, No. 101169765).

\begin{figure*}
    \centering
    \begin{mdframed}[linecolor=gray, linewidth=2.5pt, roundcorner=5pt, backgroundcolor=white] 
   \includegraphics[width=\linewidth]{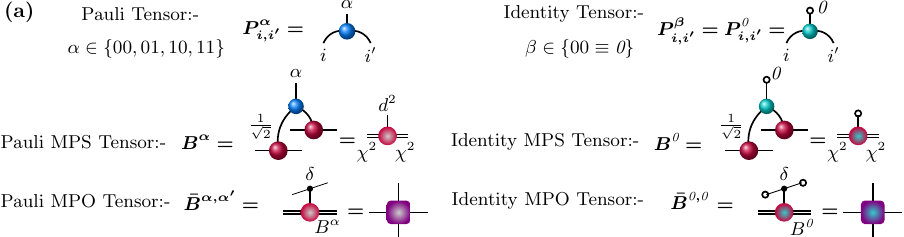}
   \end{mdframed}
   \vspace{-10pt}
    \begin{mdframed}[linecolor=gray, linewidth=2.5pt, roundcorner=5pt, backgroundcolor=white] 
  \hspace{-8pt}
\includegraphics[width=0.485\linewidth]{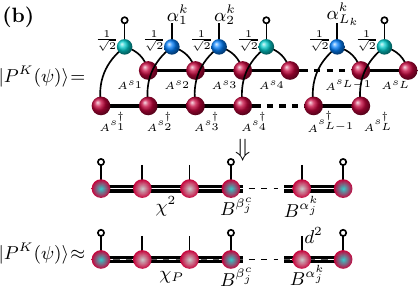}
\hspace{25pt}
  \includegraphics[width=0.47\linewidth]{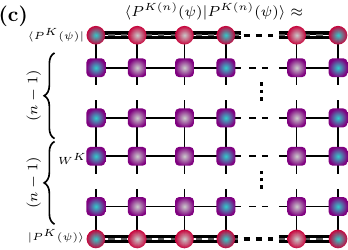}
  \end{mdframed}
    \caption{(a) Tensor-network representation of the various Pauli and identity tensors, along with the MPS and MPO structures used in the computation of the SRE. (b) Construction of the Pauli vector/MPS ($|P^K(\psi)\rangle$), obtained by contracting the Pauli tensors  ($P^\alpha_{i,i^\prime}$) with the state MPS ($|\psi\rangle$) for a subsystem $K$ of interest, and the identity tensors, $P^\beta_{i,i^\prime}=P^\textit{0}_{i,i^\prime}$, for the rest of the system, followed by truncation to a suitable bond dimension ($\chi_P$), (c) Schematic illustration of the overlap of replicated Pauli MPSs, $\langle P^{K(n)}(\psi)|P^{K(n)}(\psi)\rangle$, up to replica order $n\geq 1$, 
    which is subsequently used to compute the SRE of any partition $K$ by substituting into Eq.~\ref{eq:sremix_replica}.}
    \label{fig:SRE_PMPS}
\end{figure*}

\section{\label{sec:method}Methods}
To compute the stabilizer Rényi entropy (SRE) in different partitions, we present a complementary approach based on the Pauli-MPS. First, we introduce the Pauli-MPS algorithm based on replica tensor network contraction, which enables efficient evaluation of the SRE for many-body states represented as matrix product states. More concretely, we represent the Pauli MPS in Fig.~\ref{fig:SRE_PMPS} (a)-(b). We
express it as a two-dimensional tensor network, as shown in Fig.~\ref{fig:SRE_PMPS} (c), thereby enabling approximate contraction
using established MPS methods.
Second, we extend this method to calculate the SRE in different partitions, which allows unbiased estimation of topological nonstabilizerness. These tools provide robust and scalable access to SRE for different partitions in large-scale many-body simulations.

\textit{Pauli MPS:} We numerically calculate the full-state SRE using the replica MPS.
In particular, to calculate the SRE, we used the replica Pauli MPS method developed in Ref.~\cite{tarabunga24a}. This method begins with an  MPS representation of the state $|\psi\rangle$.
In Fig. \ref{fig:SRE_PMPS}, we present the algorithms to compute the SRE of order $n>1$ through a 2D TN. In particular, in Fig.~\ref{fig:SRE_PMPS} (b), we show how to construct the Pauli-MPS. This allows one to compute not only the SRE.

Let us consider a system of $L$ qubits in a pure state $|\psi \rangle$ given by an MPS of bond dimension $\chi$:
\begin{equation} \label{eq:mps}
    |\psi \rangle=\sum_{s_1,s_2,\cdots,s_L} A^{s_1}_1 A^{s_2}_2 \cdots A^{s_L}_L |s_1,s_2,\cdots s_L \rangle
\end{equation}
with $A_i^{s_i}$ being $\chi$\,$\times$\,$\chi$ matrices, except at the left (right)
boundary where $A_1^{s_1}$ ($A_L^{s_L}$) is a $1$\,$\times$\,$\chi$ ($\chi$\,$\times$\,$1$) row (column) vector. Here $s_i$\,$\in$\,$\left \lbrace 0, 1 \right \rbrace$  is a local computational basis.
The state is assumed right-normalised, namely $\sum_{s_i} A_i^{s_i \dagger} A_i^{s_i}$\,$=$\,$1$.

Let us define the binary string $\boldsymbol{\alpha}$\,$=$\,$(\alpha_1, \cdots, \alpha_L)$ with $\alpha_j$\,$\in$\,$\{00,01,10,11\}$.
The Pauli strings are defined as $P_{\boldsymbol{\alpha}} $\,$=$\,$ P_{\alpha_1}  \otimes P_{\alpha_2} \otimes \cdots  \otimes P_{\alpha_L}$
where $P_{00}=I, P_{01}$\,$=$\,$X, P_{11}$\,$=$\,$Y,$ and $P_{10}$\,$=$\,$Z$. We define the Pauli vector of $|\psi \rangle$ as $|P(\psi) \rangle$ with elements $  \langle \boldsymbol{\alpha} |P(\psi) \rangle$\,$=$\,$ \langle \psi | P_{\boldsymbol{\alpha}} |\psi \rangle / \sqrt{2^L}$.  When $|\psi \rangle$ has an MPS structure as in Eq. \eqref{eq:mps}, the Pauli vector can also be expressed as an MPS as follows
\begin{equation} \label{eq:PauliMPS}
    |P(\psi) \rangle=\sum_{\alpha_1,\alpha_2,\cdots,\alpha_N} B^{\alpha_1}_1 B^{\alpha_2}_2 \cdots B^{\alpha_L}_L | \alpha_1, \cdots, \alpha_L \rangle,
\end{equation}
where $B^{\alpha_i}_i $\,$=$\,$ \sum_{s,s'} \langle s | P_{\alpha_i} | s' \rangle A^s_i \otimes \overline{A^{s'}_i} / \sqrt{2}$ are $\chi^2 \times \chi^2$ matrices. Note that the MPS is normalized due to the relation $\frac{1}{2^L} \sum_{\boldsymbol{\alpha}} \langle \psi | P_{\boldsymbol{\alpha}} | \psi \rangle^2 $\,$=$\,$ 1$ which holds for pure states.
To calculate the SREs, we define a diagonal operator $W$ whose diagonal elements are the components of the Pauli vector, $\langle  \boldsymbol{\alpha'}| W |\boldsymbol{\alpha} \rangle= \delta_{\boldsymbol{\alpha'},\boldsymbol{\alpha}} \langle \boldsymbol{\alpha'} |P(\psi) \rangle$. The MPO form of $W$ reads
\begin{align}
    W =\sum_{\boldsymbol{\alpha},\boldsymbol{\alpha'}} \overline{B}^{\alpha_1,\alpha'_1}_1 \overline{B}^{\alpha_2,\alpha'_2}_2 &\cdots \overline{B}^{\alpha_L,\alpha'_L}_L\nonumber\\ &| \alpha_1, \cdots, \alpha_L \rangle \langle \alpha'_1, \cdots, \alpha'_L |,
\end{align}
where $\overline{B}^{\alpha_i,\alpha'_i}_i$\,$=$\,$ B^{\alpha_i}_i \delta_{\alpha_i,\alpha'_i}$.
Applying $W$, $n$\,$-$\,$1$ times to $|P(\psi) \rangle$, we obtain $|P^{(n)}(\psi) \rangle $\,$=$\,$W^{n-1} |P(\psi) \rangle$, which is a vector with elements $\langle \boldsymbol{\alpha} |P^{(n)}(\psi) \rangle$\,$=$\,$\langle \psi | P_{\boldsymbol{\alpha}} |\psi \rangle^n / \sqrt{2^{Ln}}$. 
We denote the local tensors of $|P^{(n)}(\psi) \rangle$ by $B^{(n)\alpha_i}_i$.
We have $\sum_{\boldsymbol{\alpha}} \langle \psi | P_{\boldsymbol{\alpha}} | \psi \rangle^{2n}/2^{Ln} = \langle P^{(n)}(\psi)  | P^{(n)}(\psi)  \rangle$ 
and the SRE (Eq.~\ref{eq:sre1}) for order $n>1$ as
\begin{equation} \label{eq:sre_replica}
    M_n(|\psi\rangle) = \frac{1}{1-n} \log_2{\langle P^{(n)}(\psi)  | P^{(n)}(\psi)  \rangle} - L\log_2 2.
\end{equation}
The Pauli-MPS itself can also be approximated with a bond dimension $\chi_P$\,$<$\,$\chi^2$. The computational cost of this compression is $O\left( \chi_P^2 \chi^2 + \chi^3 \chi_P \right)$. This is particularly advantageous for states with exponentially decaying entanglement spectra (e.g., gapped phases), in which case $\chi_P$ can be truncated to a value much smaller than $\chi^2$.

\textit{Partitioned Pauli MPS:}
Similar to the full-state SRE, we can compute the SRE of any subsystem using the replica MPS approach and using the definition of subsystem SRE as given in Eq.~\ref{Mix_SRE}. For a subsystem of interest, the subsystem SRE takes the form
\begin{align}
\tilde{M}^K_n(|\psi\rangle)=\frac{1}{1-n}\log_2\Bigg\{\frac{\displaystyle\sum_{P\in\mathcal{P}^\prime_{L_K}} |\langle\psi| P|\psi\rangle|^{2n}}{\displaystyle\sum_{P\in\mathcal{P}^\prime_{L_K}}|\langle\psi| P|\psi\rangle|^2}\Bigg\},
\label{Mix_SRE_psi}
\end{align}
where $|\psi\rangle$ is the full state represented as an MPS, and the sum runs over the restricted Pauli group $\mathcal{P}^\prime_{L_K}$ acting only on the subsystem $K$ (with identity operators on the complementary part).
We can explicitly specify a binary string $\boldsymbol{\alpha^k} = (\alpha^k_1, \cdots, \alpha^k_{L_K})$, with $\alpha^k_j \in {00, 01, 10, 11}$ corresponding to the local Pauli operator on site $j$ over the subsystem $K$ of length $L_K$. The binary string over the complementary subsystem is represented by $\boldsymbol{\beta^c} = (\beta^c_1, \cdots, \beta^c_{L-L_K})$, where $\beta^c_j = 00 \equiv \textit{0}$ ensures that all operators outside $K$ act as identities, as illustrated in Fig.~\ref{fig:SRE_PMPS}(a).
With these definitions, the modified Pauli strings take the form $P_{\boldsymbol{\alpha^k}} $\,$=$\,$P_{\textit{0}}\otimes P_{\alpha^k_1}\otimes P_{\alpha^k_2}  \otimes P_{\textit{0}}\otimes \cdots  \otimes P_{\alpha^k_{L_K}} \otimes P_{\textit{0}}$, 
where $P_{00}$\,$=P_{\textit{0}}=$\,$I, P_{01}$\,$=$\,$X, P_{11}$\,$=$\,$Y,$ and $P_{10}$\,$=$\,$Z$.  Note that the tensor-product order depends on the site index of the chain/system. 

We then define the Pauli vector of $|\psi\rangle$ for subsystem $K$ as $|P^K(\psi)\rangle$, with components $\langle \boldsymbol{\alpha^k} |P^K(\psi) \rangle$\,$=$\,$ \langle \psi | P_{\boldsymbol{\alpha^k}} |\psi \rangle / \sqrt{2^L}$ as represented in Fig.~\ref{fig:SRE_PMPS}(b).  When $|\psi \rangle$ has an MPS form as in Eq.~\eqref{eq:mps}, the Pauli vector can itself be expressed as an MPS
\begin{align} \label{eq:subPauliMPS}
    |P^K(\psi) \rangle=\sum_{\substack{\alpha^k_1,\alpha^k_2,\\\cdots,\alpha^k_{L_K}}}B^{\textit{0}}_1 B^{\alpha^k_1}_2 B^{\alpha^k_2}_3B^{\textit{0}}_4\cdots B^{\alpha^k_{L_K}}_{L-1} B^{\textit{0}}_L |\boldsymbol{\alpha^k\rangle},
\end{align}
where  $B^{\alpha^k_j}_i $\,$=$\,$ \sum_{s,s'} \langle s | P_{\alpha_i} | s' \rangle A^s_i \otimes \overline{A^{s'}_i} / \sqrt{2}$, $B^{\textit{0}}_i $\,$=$\,$ \sum_{s} A^s_i \otimes \overline{A^{s}_i} / \sqrt{2}$ are $\chi^2 \times \chi^2$ matrices and $\boldsymbol{|\alpha^k\rangle}=|\textit{0},\alpha^k_1,\alpha^k_2,\textit{0}, \cdots,\alpha^k_{L_K},\textit{0} \rangle$. The normalization of the Pauli vector is given by 
$\frac{1}{2^L} \big(\sum_{\boldsymbol{\alpha^k}} \langle \psi | P_{\boldsymbol{\alpha^k}} | \psi \rangle^2 $, which for the full system reduces to the standard normalization of a pure state, i.e., $\frac{1}{2^L} \sum_{\boldsymbol{\alpha}} \langle \psi | P_{\boldsymbol{\alpha}} | \psi \rangle^2=1$

Again to calculate the SRE of this subsystem, we define a diagonal operator $W^K$ (Fig.~\ref{fig:SRE_PMPS}(a)) whose diagonal elements are the components of the modified Pauli vector, $\langle  \boldsymbol{\alpha^{k'}}| W^K |\boldsymbol{\alpha^k} \rangle= \delta_{\boldsymbol{\alpha^{k'}},\boldsymbol{\alpha^k}} \langle \boldsymbol{\alpha^{k'}} |P^K(\psi) \rangle$. The MPO form of $W^K$ reads
\begin{align}
    W^K=\sum_{\boldsymbol{\alpha^k},\boldsymbol{{\alpha^{k}}^\prime}}\overline{B}^{\textit{0},\textit{0}}_1\overline{B}^{\alpha^k_1,\alpha^{k^\prime}_1}_2 \cdots \overline{B}^{\alpha^k_{L_K},\alpha^{k^\prime}_{L_K}}_{L-1} \overline{B}^{\textit{0},\textit{0}}_L \boldsymbol{|\alpha^k\rangle\langle\alpha^{k^\prime}|},
\end{align}
where $\boldsymbol{|\alpha^k\rangle}=|\textit{0},\alpha^k_1,\alpha^k_2,\textit{0}, \cdots,\alpha^k_{L_K},\textit{0} \rangle$ and $\boldsymbol{\langle\alpha^{k^\prime}|}=\langle\textit{0},\alpha^{k^\prime}_1,\alpha^{k^\prime}_2,\textit{0}, \cdots,\alpha^{k^\prime}_{L_K},\textit{0}|$. The other components are  $\overline{B}^{\alpha^k_j,\alpha^{k^\prime}_j}_i$\,$=$\,$ B^{\alpha^k_j}_i \delta_{\alpha^k_j,\alpha^{k^\prime}_j}$ for $\alpha^k_j,\alpha^{k^\prime}_j\in\{00,01,10,11\;\forall j\}$ and $\overline{B}^{\textit{0},\textit{0}}_i$\,$=$\,$ B^{\textit{0}}_i$. 
Applying the replica operator $W^K$ a total of $n-1$ times to the Pauli vector $|P^K(\psi)\rangle$, we obtain $|P^{K(n)}(\psi) \rangle $\,$=$\,$(W^K)^{n-1} |P^K(\psi) \rangle$, which has components $\langle \boldsymbol{\alpha^k} |P^{K(n)}(\psi) \rangle$\,$=$\,$\langle \psi | P^K_{\boldsymbol{\alpha^k}} |\psi \rangle^n / \sqrt{2^{Ln}}$.
We denote the local tensors of $|P^{K(n)}(\psi)\rangle$ by $B^{(n)\alpha^k_j}_i$ when site $i$ belongs to the subsystem $K$, and by $B^{(n)\textit{0}}_i$ when it belongs to the complement. The overlap of this replicated Pauli vector is shown in Fig.~\ref{fig:SRE_PMPS}(c) gives $\sum_{\boldsymbol{\alpha^k}} \langle \psi | P^K_{\boldsymbol{\alpha^k}} | \psi \rangle^{2n}/2^{Ln} = \langle P^{K(n)}(\psi)  | P^{K(n)}(\psi)  \rangle$.  
This leads to the expression for the subsystem stabilizer Rényi entropy (SRE) of order $n>1$ given in Eq.~\ref{Mix_SRE_psi} as 
\begin{align} \label{eq:sremix_replica}
    \tilde{M}^K_n (|\psi\rangle)= \frac{1}{1-n} \log_2{\langle P^{K(n)}(\psi)  | P^{K(n)}(\psi)  \rangle} -L\log_2 2\nonumber\\- \frac{1}{1-n}\log_2{\langle P^{K}(\psi)  | P^{K}(\psi)  \rangle}.
\end{align}
Clearly, for full pure state last term vanishes as $\sum_{\boldsymbol{\alpha^k}} \langle \psi | P^K_{\boldsymbol{\alpha^k}} | \psi \rangle^{2}/2^{L}=\langle P^{K}(\psi)  | P^{K}(\psi)  \rangle=1$ and above relation of SRE becomes the Eq.~\ref{eq:sre_replica}.

\bibliography{biblio}

\clearpage
\onecolumngrid
\appendix

\section{\label{sec:TEE}Topological Entanglement Entropy (TEE) }
We consider a one-dimensional (1D) chain described by a local Hamiltonian, where the ground state is known to exhibit both trivial and non-trivial quantum orders. These orders can be captured by computing the topological entanglement entropy (TEE) using a disconnected partition, which is ring-shaped in the case of periodic boundary conditions~\cite{zeng2015,Zeng2016,Fromholz2020,Nehra2025controlling}. In this quadri-partition setting, the TEE is defined as
\begin{align}
    S^{q}_{topo}=S_{AB}+S_{BC}-S_B-S_{ABC},
\end{align}
where $S_{AB}$, $S_{BC}$, $S_{B}$, $S_{ABC}$ represent the entanglement entropies for the respective partitions as shown in Fig.~\ref{fig:intro}(b). This formulation cancels out the area-law terms and other non-universal finite-size contributions in 1D chains, leaving only the universal contribution. A non-zero value of $S^q_{topo}$ indicates non-trivial topological order.
\begin{figure}[h!]
    \centering
    \includegraphics[width=0.7\linewidth]{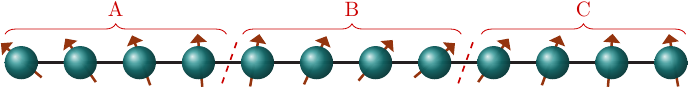}
    \caption{The spin-chain divided in three partition such that $A=[1,L/3]$, $B=(L/3,2L/3]$ and $C=(2L/3,L]$.}
    \label{fig:EEtri}
\end{figure}
An alternative partitioning scheme of interest is the connected tripartition, for which the topological entanglement entropy is defined as
\begin{align}
    S^{t}_{topo}=S_{AB}+S_{BC}-S_B-S_{ABC},
    \label{eq:TEEtri}
\end{align}
as shown in Fig.~\ref{fig:EEtri}. In this construction, the contribution from a tripartition reflects a trivial order or quantum phase transition as for pure state $S_{ABC}$ becomes zero and effectively captures the mutual information or long-range entanglement between two connected segments $AB,BC$. If a phase transition is topological, both the quadri-partition and tripartition will capture non-zero universal values for the entanglement entropy. However, in the case of a trivial phase, only the tripartition captures the phase transition, while the quadri-partition shows no transition, reflecting the absence of non-trivial topological order.

\begin{figure}[h!]
\centering
\includegraphics[width=0.32975\linewidth]{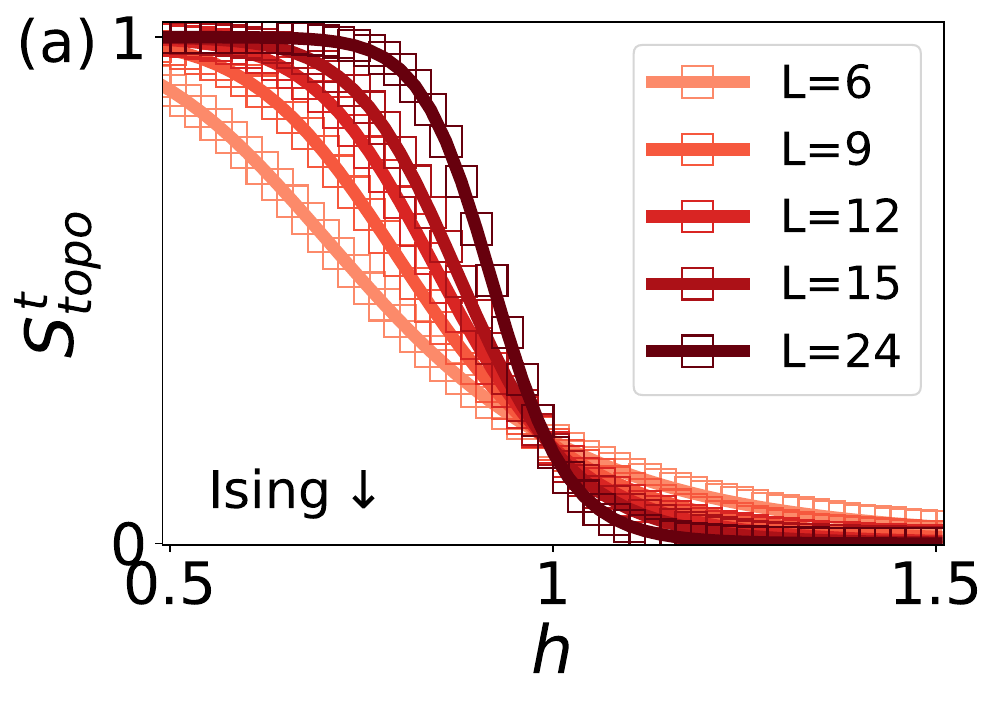}
\includegraphics[width=0.32975\linewidth]{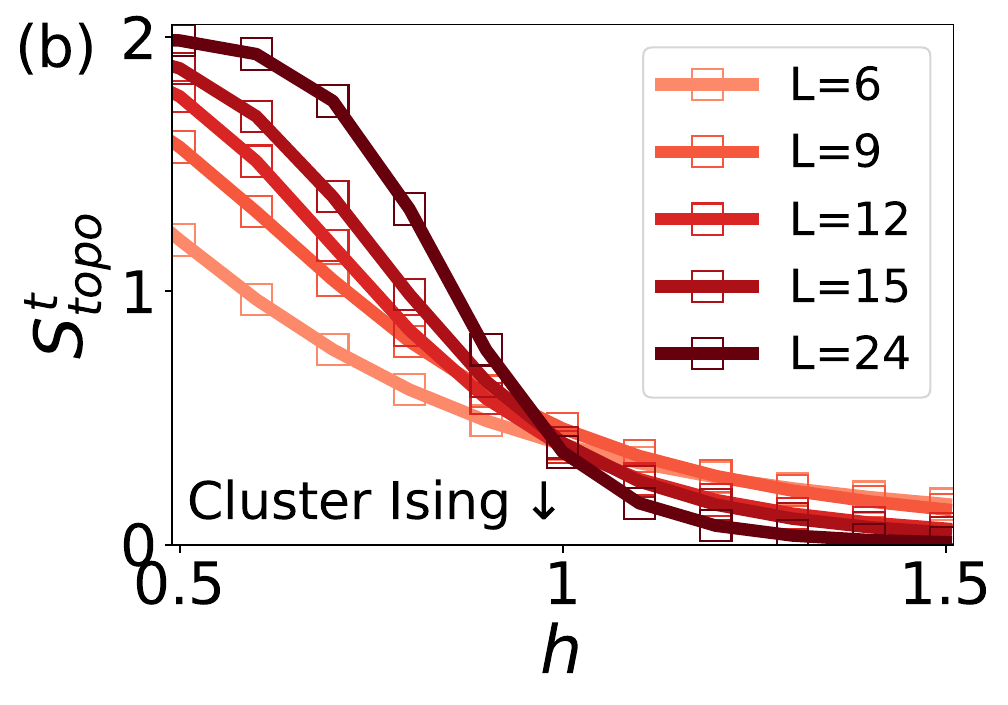}
\includegraphics[width=0.32975\linewidth]{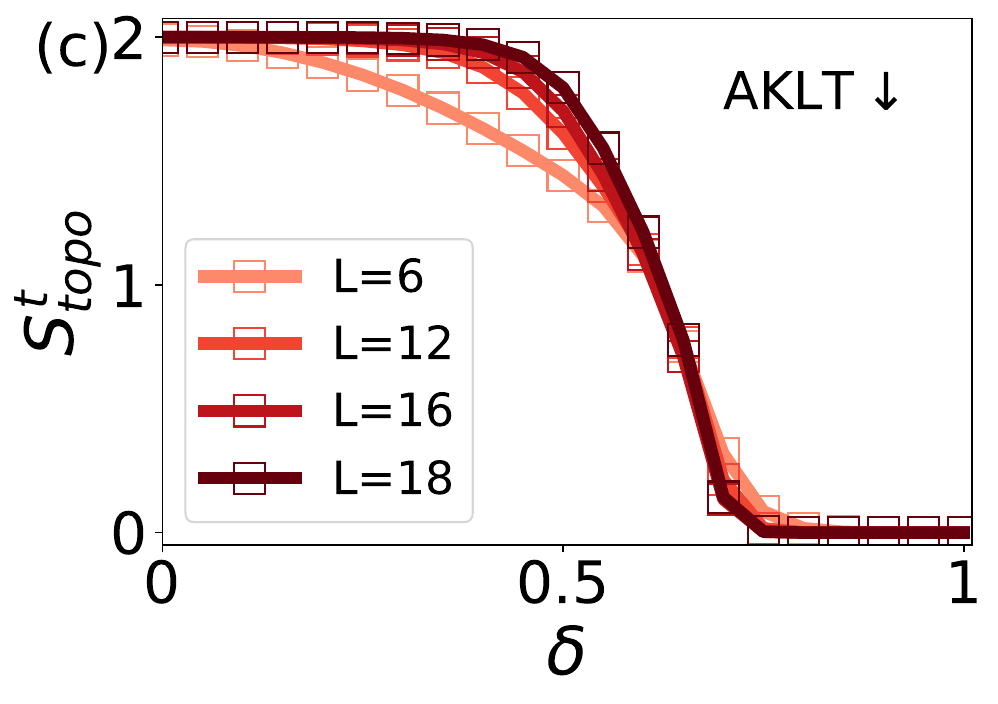}\\
\includegraphics[width=0.32975\linewidth]{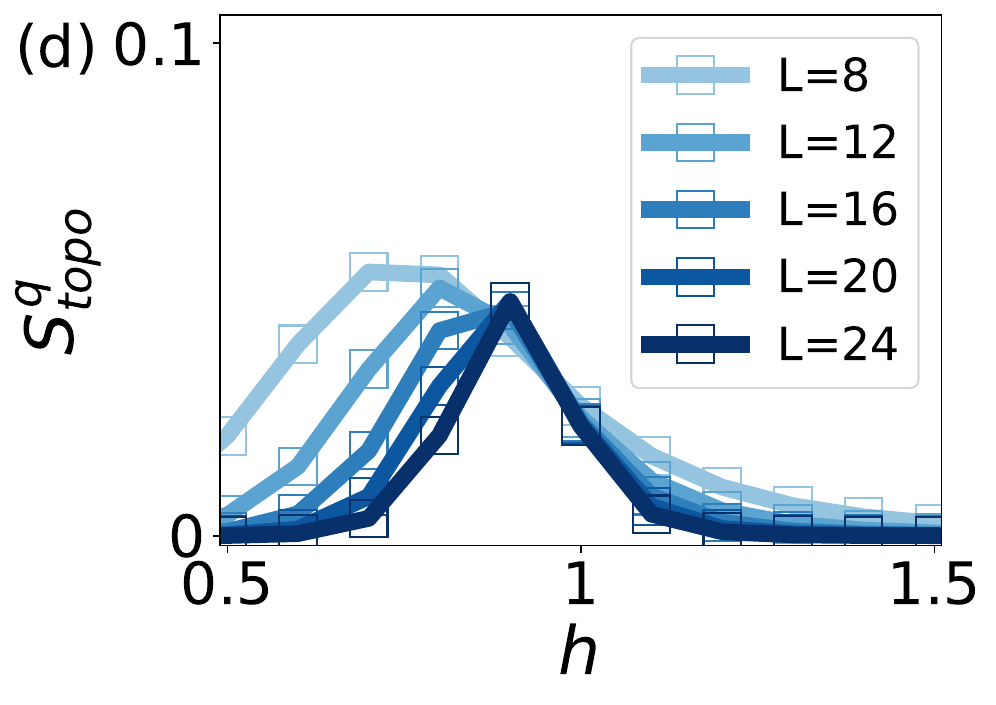}
\includegraphics[width=0.32975\linewidth]{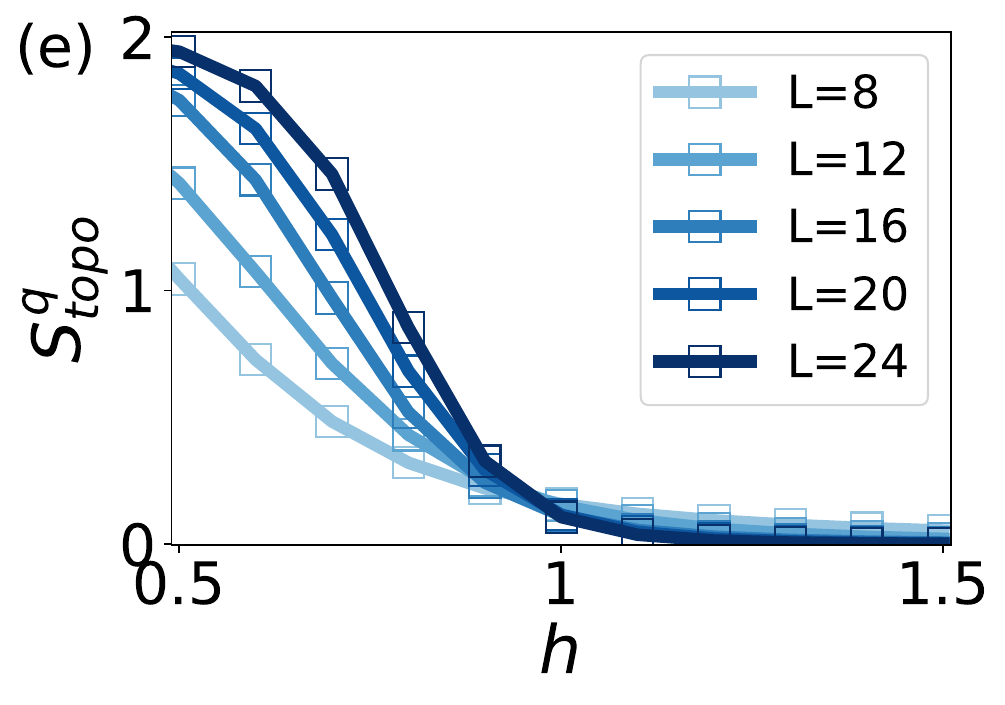}
\includegraphics[width=0.32975\linewidth]{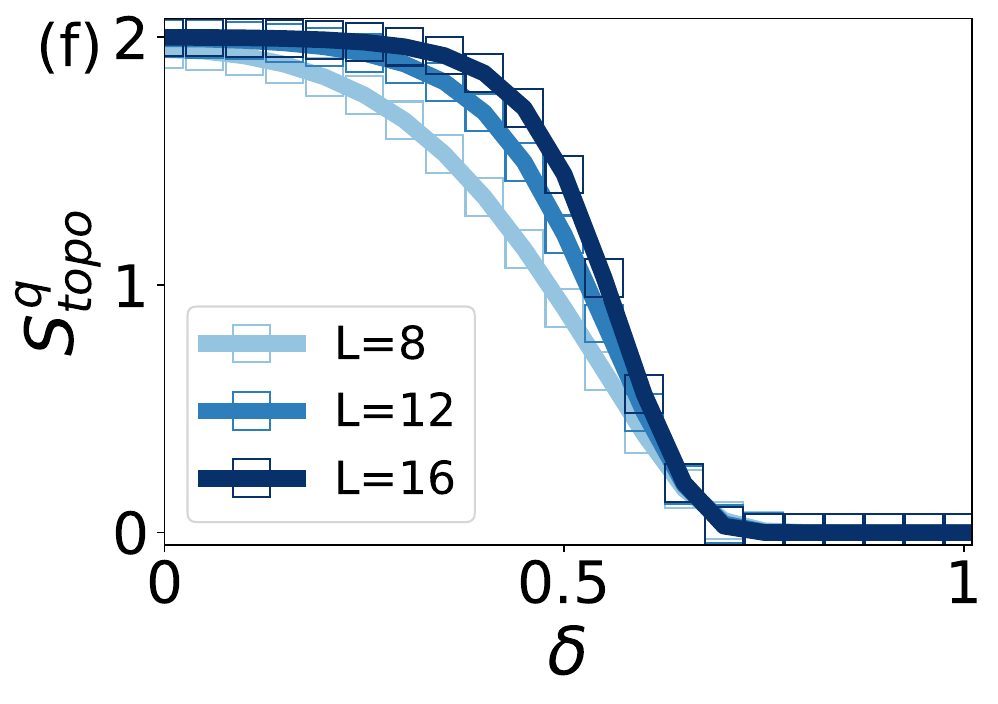}
\caption{The tri topo EE ($S^t_{topo}$) and quadri topo EE ($S^q_{topo}$) of the ground states of (a,d) Ising and (b,e) Cluster Ising model as a function of transverse field strength ($h$) and (c,f) AKLT model as a function of tuning parameter $\delta$.}
\label{fig:topoEE}
\end{figure}
As illustrative examples, we consider the ground states of local spin-1/2 Hamiltonians, specifically, the transverse-field Ising model and the Cluster-Ising model, as functions of the transverse-field strength $h$. In the Ising model, which shows a trivial (non-topological) phase transition from a ferromagnetic to a paramagnetic phase, the quadripartition TEE remains close to zero ($S_{topo}^q\approx0$) across all $h$, showing no signature of topological order in Fig.~\ref{fig:topoEE}(d).
In contrast, the tripartition TEE, $S_{topo}^t$, clearly signals the phase transition in  Fig.~\ref{fig:topoEE}(a), indicating a change from long-range correlations in the ferromagnetic regime to short-range correlations in the paramagnetic regime, where mutual information between different regions cancels out (see also~\cite{tarabunga2025bell}).
However, the Cluster-Ising model hosts a topological quantum phase transition from a non-trivial Cluster phase to a trivial paramagnetic phase.
Here, both $S^t_{\mathrm{topo}}$ and $S^q_{\mathrm{topo}}$ detect the transition in Fig.~\ref{fig:topoEE}(b,e), but the quadripartition TEE remains finite in the cluster phase, reflecting the presence of non-trivial topological order. This nonzero TEE originates from the long-range entanglement of edge modes in the cluster phase, even though all local correlations decay exponentially. 

We also extend this analysis to spin-1 models, focusing on both the tripartition and quadripartition TEEs, $S^t_{topo}$, $S^q_{topo}$. Here, we examine the one-parameter family of 1D spin-1 Hamiltonians defined in Eq.~\ref{eq:hamAKLT}~\cite{PhysRevLett.121.140604}, which provides a rich platform to explore topological phase transitions in higher-spin systems. As the parameter $\delta$ 
varies from 0 to 1, the system undergoes a topological phase transition from the Haldane phase, which exhibits non-trivial topological order at small $\delta$, to a trivial phase for $\delta$ close to 1. 
The Haldane phase is a symmetry-protected topological (SPT) phase, characterized by the presence of robust edge modes and protected by $\mathrm{Z}_2\times \mathrm{Z}_2$ symmetry (i.e., the $\pi$ rotations about the $x$ and $z$ axes) and bond-centered reflection symmetry. 
The long-range entanglement nature of edge states in this phase is captured by both $S^t_{topo}$, $S^q_{topo}$, which remain finite in the Haldane phase and vanish in the trivial phase. This behavior is similar to what is observed in the Cluster-Ising spin-1/2 model, showing that TEEs can effectively reveal topological order in these cases.

\section{Tri-partition SRE ($M^t_{topo}$)}
As discussed in the previous section, both tripartition and quadripartition entanglement entropies capture universal signatures of topological phases. A nonzero contribution in both indicates a topological phase, while a nonzero contribution appearing only in the tripartition TEE signals trivial long-range entanglement. The vanishing of both corresponds to a fully trivial phase without long-range entanglement. In the main text, we only focused on the quadri-partition topological stabilizer Rényi entropy (SRE), which reveals nontrivial magic in the topological phases of various models. It is therefore natural and instructive to extend this analysis to the tripartition case. 
Accordingly, in this section, we study the tripartition topological SRE,
\begin{align}
    M^t_{topo}=-(\tilde{M}_{AB}+\tilde{M}_{BC}-\tilde{M}_{B}-\tilde{M}_{ABC})
    \label{eq:triSRE}
\end{align}
defined in direct analogy with the tripartition TEE ($S^t_{topo}$) in Eq.~\ref{eq:TEEtri} for the partitions $A=[1,L/3]$, $B=(L/3,2L/3]$, and $C=(2L/3,L]$ as shown in Fig.~\ref{fig:EEtri}. 

\begin{figure}[h!]
\centering
\includegraphics[width=0.323\linewidth]{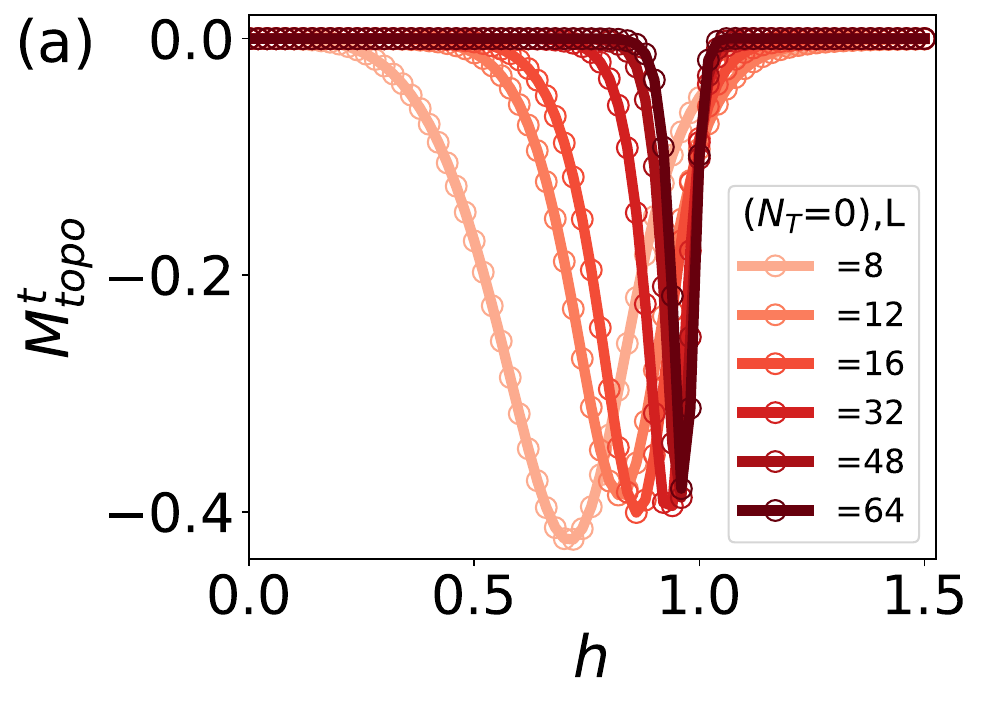}
\includegraphics[width=0.323\linewidth]{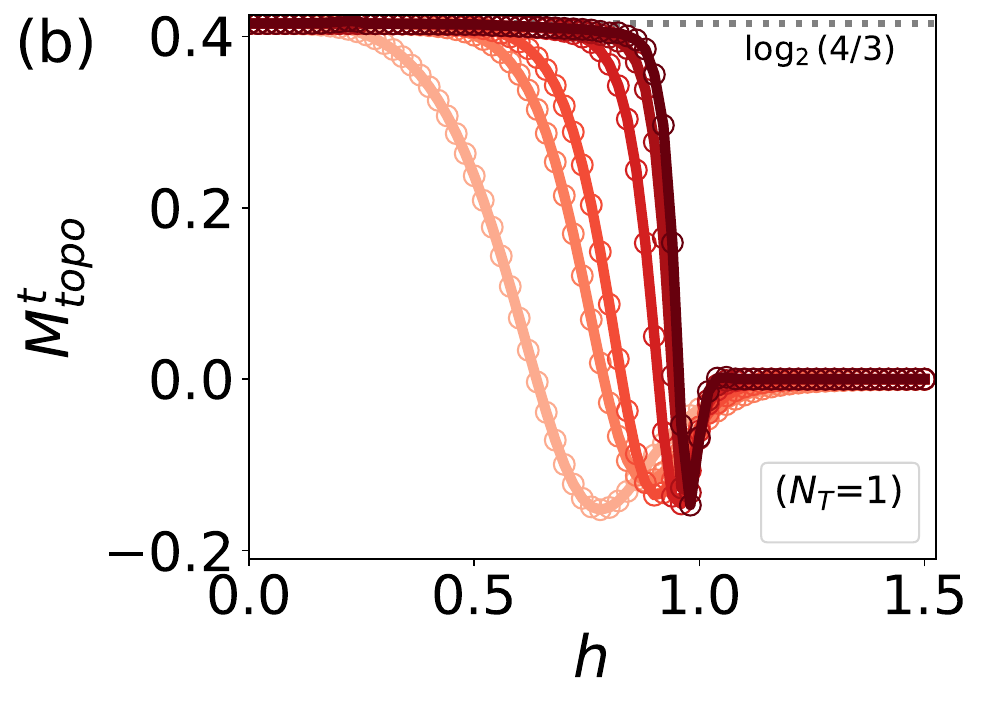}
\includegraphics[width=0.323\linewidth]{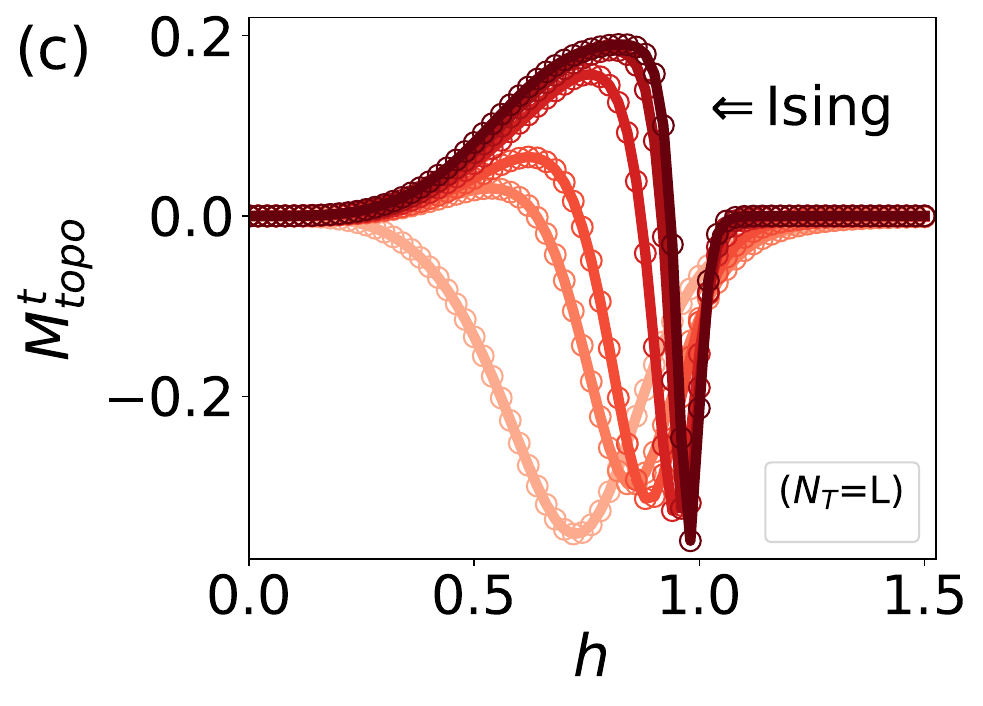}\\
\includegraphics[width=0.323\linewidth]{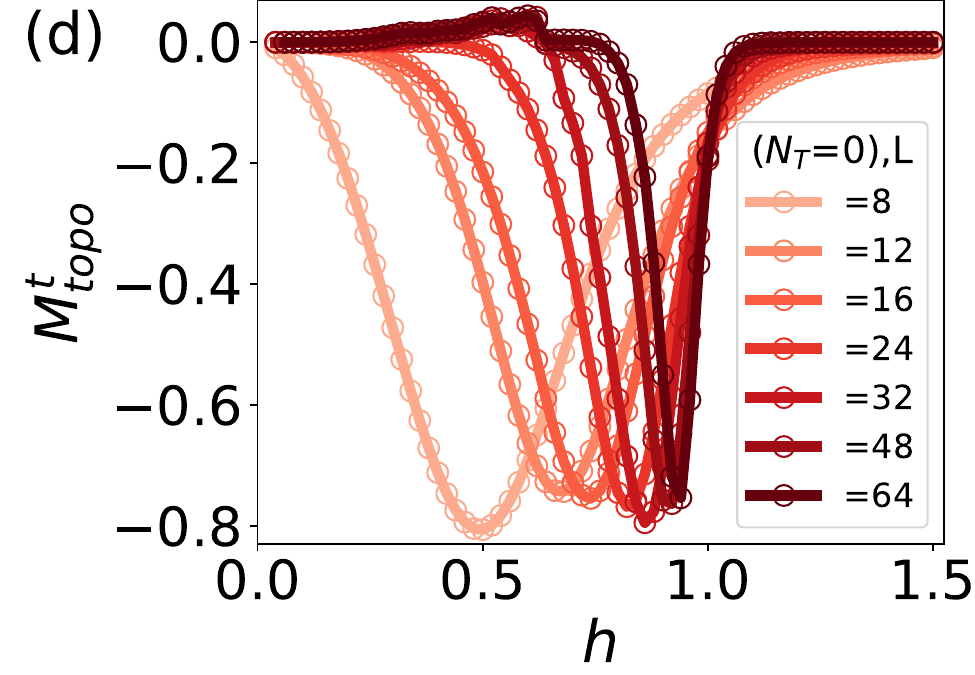}
\includegraphics[width=0.323\linewidth]{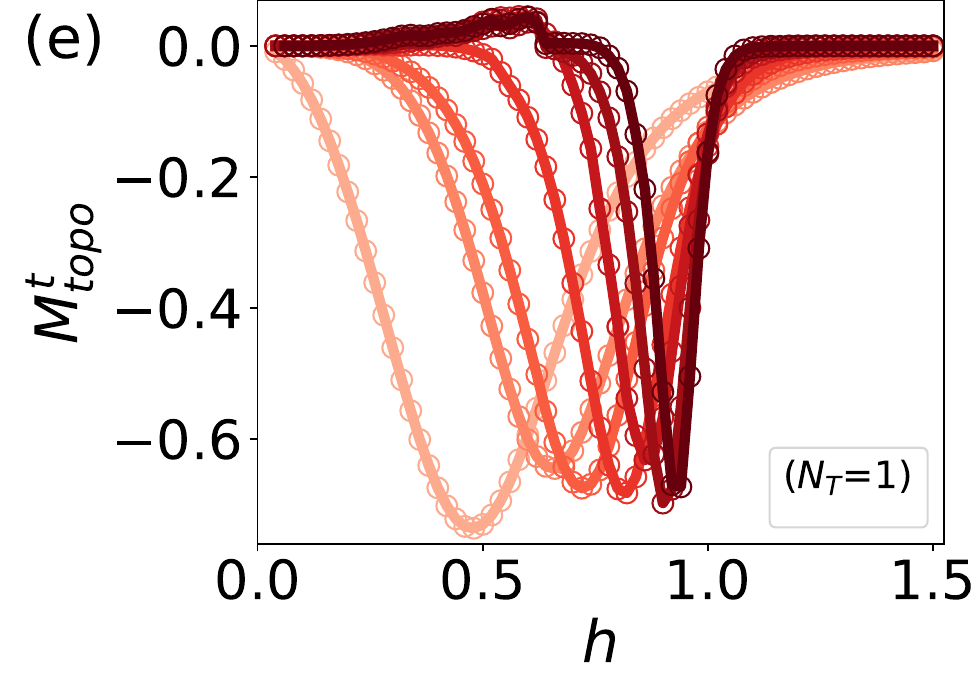}
\includegraphics[width=0.323\linewidth]{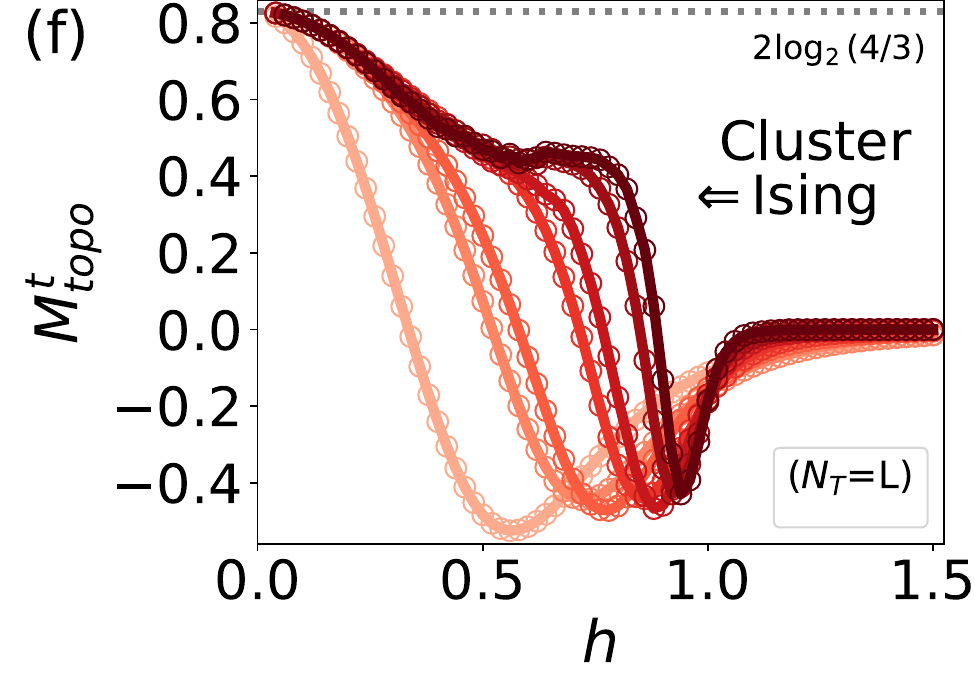}
\caption{Tri-partition topological SRE ($M^t_{topo}$) of the (a–c) Ising model and (d–f) Cluster Ising model with (a,d) $N_T=0$, (b,e) $N_T=1$, and (c,f) $N_T=L$ T-gates applied to the ground state of a chain of length $L$, as a function of transverse field $h$ with $J=1$.}
\label{fig:tri_Magic}
\end{figure}
We now present results for the tri-partition topological SRE ($M^t_{topo}$) obtained using the Pauli MPS method for spin-1/2 and spin-1 models. We begin with the transverse-field Ising model, which undergoes a trivial symmetry-protected phase transition that has also been studied using topological entanglement entropy (Figs.~\ref{fig:topoEE}(a,b)).
It is well known that the full-state magic captures the Ising phase transition: it peaks near the critical point, with the peak height scaling with system size, and vanishes deep in both phases where the ground state reduces to a stabilizer state~\cite{tarabunga23m}. Partitioned magic exhibits similar behavior, with peak heights depending on the subsystem size. In the tri-partition setup, for example, partitions AB and BC yield identical behavior due to equal length, while the full chain (ABC) exhibits the largest peak. The peak position is also shaped by boundary effects: bulk partitions effectively mimic periodic boundary conditions and peak closer to the critical point (e.g., partition B), while boundary partitions deviate slightly. As a result, the combined measure defined in Eq.~\ref{topomagic} for three cuts Fig.~\ref{fig:intro}(b)) shows a dip at the transition for finite systems, vanishing in both phases due to stabilizer character as shown in Fig.~\ref{fig:tri_Magic}(a). In the thermodynamic limit, all partition peaks align at the critical point $h_c=1.0$, and $M^t_{topo}$ develops a sharp dip exactly at the transition while remaining zero elsewhere, providing a clean diagnostic of the phase boundary. This behavior is also confirmed by analytical results presented in Sec.~\ref{sec:anal}. Since all subsystem contributions to the SRE vanish in the tripartition scheme, the entire topological SRE originates from the global state. In this case, $M^t_{\rm topo} = \tilde{M}^{ABC}$ with $\tilde{M}^{AB} = \tilde{M}^{BC} = \tilde{M}^{B} = 0$, indicating that the GHZ state hosts only global nonstabilizerness without any local contributions.

In contrast, when a single T-gate is applied, the Ferromagnetic-ordered regime loses its stabilizer character and contributes a finite amount of magic equal to that of a single T-gate, i.e., $\log_2(4/3)$ for the full chain. Because this regime is highly symmetric and exhibits long-range correlations, the magic of individual partitions increases slightly, but full state magic increases by a T-gate contribution, so the net difference is still $\log_2(4/3)$. In the paramagnetic regime, the magic contributions scale with partition length and cancel, causing $M^t_{topo}$ again vanishes. Thus, tri-partition magic provides a sharp diagnostic of the transition, as seen in Fig.~\ref{fig:tri_Magic}(b), distinguishing the symmetry-ordered phase with $\log_2(4/3)$ from the disordered phase with zero, similar to the behavior of $S^t_{topo}$. However, in the presence of an even number of T-gates ($N_T=L$), due to the periodicity of the global phase $e^{iN_T\theta}$, the ground state remains close to a stabilizer state, and $M^t_{topo}$ largely vanishes in both phases. Nevertheless, near the critical point, the breakdown of stabilizer structure and various correlations in the ordered regime give rise to a small but finite $M^t_{topo}$, as shown in Fig.~\ref{fig:tri_Magic}(c), again reflecting behavior similar to $S^t_{topo}$.

We also studied the Cluster Ising model across its SPT phase transition, which, without T-gates, Fig.~\ref{fig:tri_Magic}(d), exhibits behavior similar to the Ising model. However, the effect of T-gates is more subtle; since the cluster state is topological, with non-local edge correlations, a single T-gate is insufficient to significantly probe its structure. Consequently, the tri-partition magic remains largely unchanged as shown in Fig.~\ref{fig:tri_Magic}(e). Only when the system is heavily doped with T-gates does the interplay between non-local ground-state correlations and injected magic produce non-trivial features in the topological regime, as reflected by $M^t_{topo}$ in Fig.~\ref{fig:tri_Magic}(f), closely mirroring the quadripartite results shown in Fig.~\ref{fig:quad_Magic}(f). This picture is further supported by the analytical analysis presented in Sec.~\ref{sec:anal}. There, we show that the SRE of a connected subsystem ($K$) scales as ($L_K-2$), which leads to  
\begin{align}
M^t_{topo}=[-(L_{AB}-2)-(L_{BC}-2)+(L_B-2)+L_{ABC}]\log_2(\frac{4}{3})=2\log_2(\frac{4}{3}),
\end{align}
in direct analogy with the quadripartition SRE $M^q_{topo}$ given in Eq.~\ref{eq:CluSRE}.
Overall, the behavior of topological SRE parallels that of the topological entanglement entropy in these models, as shown in the table~\ref{tabletri} below.
\begin{table}[h!]
\centering
\renewcommand{\arraystretch}{1.4} 
\begin{tabular}{|c|c|c|c|}
\hline
\rowcolor{gray!40}
\multicolumn{1}{|c|}{\parbox{4cm}{\centering \diagbox[height=2em]{\;\;\textbf{Quantity}}{\textbf{States}\;\;}}} 
&\;\;\textbf{PM}\;\; & \;\;\textbf{FM}\;\; & \;\;\textbf{CL}\;\; \\
\hline
$S^t_{topo}$ & 0 & $\log_2(2)$ & $2\log_2(2)$ \\\hline
$M^t_{topo},\ N_T=0$ & 0 & $0$ & 0 \\\hline
$M^t_{topo},\ N_T=1$ & 0 & $\log_2(4/3)$ & 0 \\\hline
$M^t_{topo},\ N_T=L$ & 0 & $\log_2(4/3);\;L=2m+1$ or $0;\;L=2m$ & $2\log_2(4/3)$ \\\hline
\end{tabular}
\caption{Similarity of topological quantities in tri-parition setup: entanglement entropy ($S^{t}_{\text{topo}}$) and magic ($M^{t}_{\text{topo}}$) for paramagnetic (PM), ferromagnetic (FM), and cluster (CL) phases in symmetry-protected trivial and topological phases, with and without T-gates where $N_T$ is the number of T-gates and $L$ is the system size.}
\label{tabletri}
\end{table}

Next, we have also considered the spin-1 model in Eq.~\ref{eq:hamAKLT}, which exhibits topological characteristics. This model undergoes a symmetry-protected topological (SPT) phase transition, which is captured by both the tripartition and quadri-partition TEEs ($S^t_{topo}, S^q_{topo}$), as shown in Fig.~\ref{fig:topoEE}(c,f), clearly indicating a topological quantum phase transition. It is therefore natural to investigate this transition using the tripartition SRE ($M^t_{topo}$) to better understand the behavior of quantum magic in this system.
\begin{figure}[h!]
    \centering
    \includegraphics[width=0.5\linewidth]{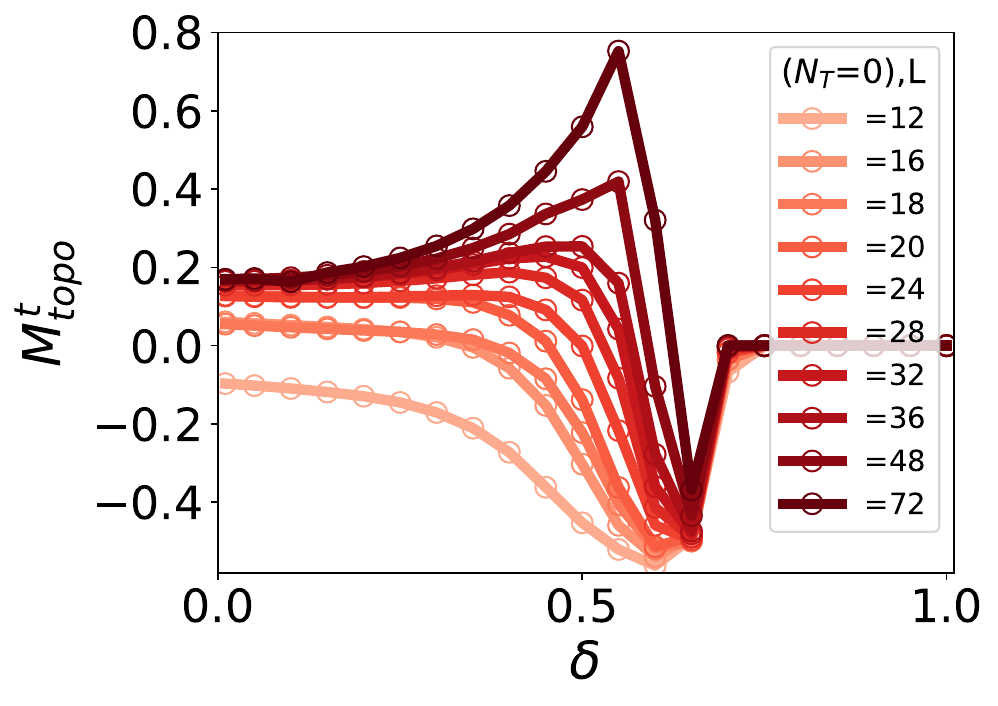}
    \caption{Tri-parititon SRE ($M^t_{topo}$) of the AKLT model (eq.~\ref{eq:hamAKLT}) as a function of tuning parameter $\delta$.}
    \label{fig:TriAKLT}
\end{figure}
Similar to the quadri-partition SRE ($M^q_{topo}$) shown in Fig.~\ref{fig:SREAKLTq}, the tripartition SRE ($M^t_{topo}$) also faithfully captures the quantum phase transition, as illustrated in Fig.~\ref{fig:TriAKLT}. It takes on a non-trivial constant value in the topological phase and drops to zero in the trivial phase. Again, the non-trivial constant value is $\approx 0.169$ in the thermodynamic limit; this saturation arises due to the constant magic contributions of the individual segments. A small difference between the bulk and boundary contributions produces this finite value and enables the tripartition SRE to distinguish the non-trivial phase of the model clearly. 

\section{Exact calculation for Tri-critical Ising model with small chain length}
Here, we examine the ground-state properties of the tri-critical Ising (TCI) model, as defined in Eq.~\ref{TCIM_ham} of the main text, for a small system size. By employing the MPS representation in Eq.~\ref{MPS_Tgate}, we can compute the exact tri- and quadri-partition magic of the doped ground state of the TCI model for a chain of length $L=8$, using the following parameters:
\begin{align}
    N_m=\frac{(1+g)^8+(1-g)^8}{2},
\end{align}
\begin{align}
\xi_1=1 + 50 g^2 + 88 g^3 + 1008 g^4 + 1400 g^5 + 4974 g^6 + 4656 g^7 + 
 8414 g^8 + 4656 g^9 + 4974 g^{10} + 1400 g^{11} + 1008 g^{12} +\nonumber\\ 88 g^{13} +  50 g^{14} + g^{16},
\end{align}
\begin{align}
\xi_2=1 +g^{32}+ 12 g^2(1+g^{28}) + 512 g^4(1+g^{24}) + 512 g^5(1+g^{22}) + 12404 g^6(1+g^{20}) + 12416 g^7(1+g^{18}) + 180748\nonumber \\ g^8(1+g^{16}) + 
 173184 g^9(1+g^{14}) + 1474140 g^{10}(1+g^{12}) + 1244160 g^{11}(1+g^{10}) + 6675904 g^{12}(1+g^{8}) + \nonumber\\4386304 g^{13}(1+g^{6}) + 15864100 g^{14}(1+g^{4}) + 7454464 g^{15}(1+g^{2}) + 20593766 g^{16} ,
\end{align}
\begin{align}
\xi_3=1 + 42 g^2 + 120 g^3 + 960 g^4 + 1432 g^5 + 4982 g^6 + 4592 g^7 + 
 8510 g^8 + 4592 g^9 + 4982 g^{10} + 1432 g^{11} + 960 g^{12} +\nonumber \\ 120 g^{13} +  42 g^{14} + g^{16},
\end{align}
\begin{align}
\xi_4=1 + g^{32} +40 g^2(1+g^{28}) + 1152 g^4(1+g^{24}) + 704 g^5(1+g^{22}) + 26312 g^6(1+g^{20}) + 27200 g^7(1+g^{18}) + 372940 \nonumber \\g^8 (1+g^{16}) + 361280 g^9(1+g^{14}) + 2860472 g^{10}(1+g^{12}) + 2354624 g^{11}(1+g^{10}) + 12046912 g^{12}(1+g^8) +\nonumber \\  7656320 g^{13}(1+g^6) + 27796312 g^{14}(1+g^4) + 13192832 g^{15}(1+g^2) +  36475110 g^{16},
\end{align}
\begin{align}
\xi_5=(1 + 280 g^4 + 5376 g^6 + 75292 g^8 + 596736 g^{10} + 2743720 g^{12} + 
  6193152 g^{14} + 7644742 g^{16} + 6193152 g^{18} +\nonumber \\ 2743720 g^{20} + 
  596736 g^{22} + 75292 g^{24} + 5376 g^{26} + 280 g^{28} + g^{32}),
\end{align}
\begin{align}
 \xi_6=1 + 24 g^2 + 64 g^3 + 612 g^4 + 1536 g^5 + 4232 g^6 + 6080 g^7 + 
 7670 g^8 + 6080 g^9 + 4232 g^{10} + 1536 g^{11} + 612 g^{12} + 64 g^{13} +
 \nonumber\\
 24 g^{14} + g^{16},
 \end{align}
 \begin{align}
\xi_7=1 +  g^{32}+ 24 g^2(1+g^{28}) + 992 g^4 (1+g^{24})+ 384 g^5(1+g^{22}) + 20104 g^6(1+g^{20}) + 20608 g^7(1+g^{18}) + 266540\nonumber\\ g^8(1+g^{16}) + 
 350848 g^9(1+g^{14}) + 2111960 g^{10}(1+g^{12}) + 2838400 g^{11}(1+g^{10}) + 9530464 g^{12}(1+g^{8}) +  \nonumber\\10729216 g^{13}(1+g^6) + 23263112 g^{14}(1+g^4) + 20139264 g^{15}(1+g^2) + 31325478 g^{16},     
 \end{align}
\begin{align}
\xi_8=1 + g^{32}+ 84 g^2(1+g^{28}) + 3104 g^4(1+g^{24}) + 512 g^5 (1+g^{22})+ 66428 g^6 (1+g^{20})+ 34432 g^7(1+g^{18}) + 
 836588 \nonumber \\g^8(1+g^{16}) + 588416 g^9(1+g^{14}) + 5723540 g^{10}(1+g^{12}) + 4027392 g^{11}(1+g^{10}) + 22005536 g^{12}(1+g^{8}) +\nonumber \\ 13395456 g^{13}(1+g^{6}) + 48727708 g^{14}(1+g^{4}) + 23896832 g^{15}(1+g^{2}) + 
 63377830 g^{16}.
\end{align}
\begin{figure}[h!]
    \centering
    \includegraphics[width=0.5\linewidth]{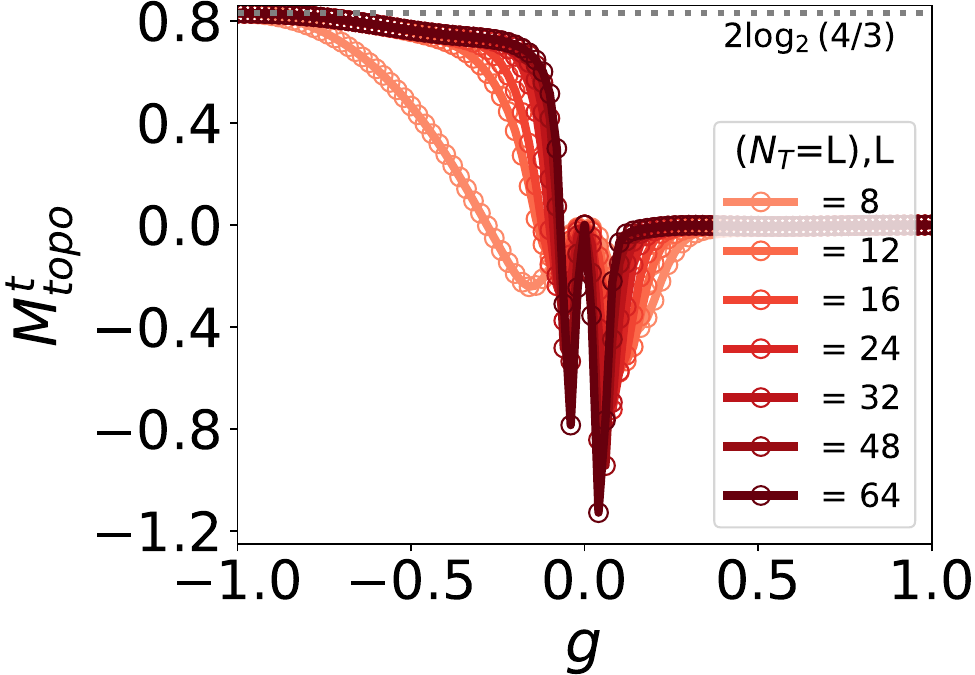}
    \caption{Tri-partition SRE ($M^t_{topo}$) of the fully T-gates doped ground state ($N_T=L$) of the Tri-critical Ising model given in eq.~\ref{MPS_Tgate} as a function of parameter $g$ for different system sizes $L$.}
    \label{fig:TCIsingtri}
\end{figure}
This leads to the quadri-partition stabilizer Rényi entropy (SRE) (Eq.~\ref{Mix_SRE}) of the chain with four cuts, as shown in Fig.~\ref{fig:intro}(b)
\begin{align}
    \tilde{M}^{AB}=\log_2\Bigg(\frac{N_m^2\xi_3}{\xi_4} \Bigg);\tilde{M}^{BC}=\log_2\Bigg(\frac{N_m^2\xi_6}{\xi_7} \Bigg);\tilde{M}^{B}=\log_2\Bigg(\frac{N_m^2\xi_1}{\xi_8} \Bigg);\tilde{M}^{ABC}=\log_2\Bigg(\frac{N_m^2\xi_1}{\xi_2} \Bigg),
\end{align}
Using Eq.~\ref{topomagic}, the quadri-partition topological SRE ($M^q_{topo}$) is then
\begin{align}
    M^q_{topo}|_{L=8}=\log_2\Bigg(\frac{\xi_3\xi_6\xi_8\xi_2}{\xi_4\xi_7\xi^2_1} \Bigg).
\label{quadmag:TCI}
\end{align}
As mentioned earlier, for particular values of $g$, the ground state reduces to well-known states, allowing for an exact computation of both SREs using Eqs.~\ref{trimag:TCI} and \ref{quadmag:TCI}. 
For $g=-1$, it becomes the Cluster state, which gives $M^q_{\rm topo}=2\log_2(4/3)$; for $g=0$ it is a GHZ state with $M^q_{\rm topo}=0$; and at $g=1$, corresponding to the paramagnetic state, topological SRE $M^q_{\rm topo}$ again vanish. These exact results are fully consistent with our numerical findings for the Ising, cluster Ising, and tri-critical Ising models presented in Figs.~\ref{fig:quad_Magic} and~\ref{fig:TCIsing}.

Similarly, the stabilizer R\'{e}nyi entropy, computed using Eq.~\ref{Mix_SRE}, for the different partition contributions under the three-cut scheme shown in Fig.~\ref{fig:EEtri}, is given by \begin{align}
    \tilde{M}^{AB}=\tilde{M}^{BC}=\log_2\Bigg(\frac{N_m^2\xi_1}{\xi_2}\Bigg);\tilde{M}^{B}=\log_2\Bigg(\frac{N_m^2\xi_3}{\xi_4}\Bigg);  \tilde{M}^{ABC}=\log_2\Bigg(\frac{N^4_m}{\xi_5}\Bigg).
\end{align}
Using Eq.~\ref{eq:triSRE}, the tri-partition topological SRE ($M^t_{topo}$) is then
\begin{align}
    M^t_{topo}|_{L=8}=\log_2\Bigg(\frac{N^2_m\xi^2_2\xi_3}{\xi_1^2\xi_4\xi_5}\Bigg).
    \label{trimag:TCI}
\end{align}
The tripartition SRE shows the same consistency across models, in agreement with the results shown in Figs.~\ref{fig:tri_Magic},~\ref{fig:TCIsingtri} for the Ising, cluster Ising, and tri-critical  Ising models. In particular, for the cluster state ($g=-1$), we find $M^t_{\rm topo}=2\log_2(4/3)$, whereas $M^t_{\rm topo}=0$ for both the GHZ state ($g=0$) and the paramagnetic state ($g=1$).
\end{document}